\documentclass[12pt]{iopart}

\usepackage{graphicx,epstopdf}  
\expandafter\let\csname equation*\endcsname\relax
\expandafter\let\csname endequation*\endcsname\relax
\usepackage{amsmath}
\usepackage{graphicx}
\usepackage{amsfonts}
\usepackage{color}
\usepackage{dcolumn}   
\usepackage{bm}        
\usepackage{amssymb}   
\usepackage{hyperref}
\usepackage{amsmath}
\usepackage{amsfonts}
\usepackage{amssymb}

\begin{document}

	\newcommand{\angstrom}{\text{\normalfont\AA}}
	\newcommand{\braket}[3]{\bra{#1}\;#2\;\ket{#3}}
	\newcommand{\projop}[2]{ \ket{#1}\bra{#2}}
	\newcommand{\ket}[1]{ |\;#1\;\rangle}
	\newcommand{\bra}[1]{ \langle\;#1\;|}
	\newcommand{\iprod}[2]{ \langle#1|#2\rangle}
	\newcommand{\intl}[2]{\int\limits_{#1}^{#2}}
	\newcommand{\logt}[1]{\log_2\left(#1\right)}
	
	\newcommand{\mc}[1]{\mathcal{#1}}
	\newcommand{\mb}[1]{\mathbb{#1}}
	\newcommand{\cx}[1]{\tilde{#1}}
	\newcommand{\nn}{\nonumber}
	\newcommand{\la}{\langle}
	\newcommand{\ra}{\rangle}
	\newcommand{\blang}{\big \langle}
	\newcommand{\brang}{\big \rangle}
	
	\newcommand{\p}{\partial}
	\newcommand{\cmnt}[2]{\textbf{\#\#}{\color{#1}#2}\textbf{\#\#}}
	
	\newcommand{\flap}{\mb{L}_{\bar{\kappa}}}
	\newcommand{\flapfinite}{\mb{L}}
	\newcommand{\fcurr}{\mb{A}}
	\newcommand{\flapFull}{|\Delta|^{3/4}}
	\newcommand{\tdir}{f}
	\newcommand{\mzeta}{\chi}
	\def\be{\begin{equation}}
	\def\ee{\end{equation}}
	\def\bea{\begin{eqnarray}}
	\def\eea{\end{eqnarray}}
	
	\newcommand{\eqa}[1]{\begin{align}#1\end{align}}
	\newcommand{\mbf}[1]{\mathbf{#1}}
	\newcommand{\iu}{{i\mkern1mu}}
	\newcommand{\widesim}[2][1.5]{
		\mathrel{\overset{#2}{\scalebox{#1}[1]{$\sim$}}}
	}
	
	\newcommand{\sina}{\alpha}
	\newcommand{\HTheta}{\theta}
	\newcommand{\genpower}{\beta}
	\newcommand{\cosa}{\phi}
	\newcommand{\RN}[1]{
		\textup{\uppercase\expandafter{\romannumeral#1}}%
	}
	
	\title[Fractional equation description of an open anomalous heat conduction set-up]{Fractional diffusion equation description of an open anomalous heat conduction set-up}

	\author[1]{Aritra Kundu$^1$, C\'edric Bernardin$^2$, Keji Saito$^3$, Anupam Kundu$^4$,  Abhishek Dhar$^5$}
	\address{$^{1,4,5}$International Centre for Theoretical Sciences, Bengaluru, India\\$^2$Universit\'e C\^ote d'Azur, CNRS, LJADNice Cedex 02, France\\$^3$Department of Physics, Keio University, Yokohama 223-8522, Japan}
	\ead{$^1$aritrak@icts.res.in,$^2$Cedric.BERNARDIN@unice.fr,$^3$saitoh@rk.phys.keio.ac.jp,\\$^4$anupam.kundu@icts.res.in,$^5$abhishek.dhar@icts.res.in,  }
	\vspace{10pt}
	\begin{indented}
		\item[]\today
	\end{indented}

	\begin{abstract}
		We provide a stochastic fractional diffusion equation description of energy transport through a finite one-dimensional chain of harmonic oscillators with stochastic momentum exchange and connected to Langevian type heat baths at the boundaries. By establishing an unambiguous finite domain representation of the associated fractional operator, we show that this equation can correctly reproduce equilibrium properties like Green-Kubo formula as well as non-equilibrium properties like the steady state temperature and current. In addition, this equation provides the exact time evolution of the temperature profile. Taking insights from the diffusive system and from numerical simulations, we pose a conjecture that these long-range correlations in the steady state are given by the inverse of the fractional operator. We also point out some interesting properties of the spectrum of the fractional operator.  All our analytical results are supplemented with  extensive numerical simulations of the microscopic system.

	\end{abstract}

	
	\section{Introduction}
	\label{intro}

Energy transport across an extended system is a fundamental non-equilibrium phenomena which is often described  by the phenomenological Fourier law. 
This law leads to the    heat equation  for the evolution of the temperature field $T(y,t)$, which in one dimension is given by  
\begin{equation}
\p_t T(y,t) = ({\kappa}/c) \p_y^2 T(y,t), \label{heat-diff}
\end{equation}
where $c$ is the specific heat capacity and ${\kappa}$ the heat conductivity (assumed, for simplicity to be  temperature independent).  This equation plays a central role in understanding heat transport through macroscopic materials in several contexts.  
However, various studies have established that for a large class of one and two dimensional  systems with momentum conservation, energy transport is not diffusive but super-diffusive --- this is referred to as anomalous transport \cite{Lepri2016a,Lepri2003,Dhar2008}.  
There are several signatures of anomalous transport which include the super-diffusive spreading of localized heat pulses and the form of  spatio-temporal correlations in equilibrium set-up as well as diverging thermal conductivity in non-equilibrium (boundary driven) set-up . 
Unlike diffusive transport, currently there are no general framework to understand these features of anomalous transport completely except for a recent development based on non-linear fluctuating hydrodynamics theory \cite{VanBeijeren2012,Spohn2014,Mendl2013,Popkov2016}. This theory, based on some phenomenological assumptions,  provides a framework to understand the super-diffusive energy transport in  a wide class of one-dimensional anharmonic classical Hamiltonian systems such as the Fermi-Pasta-Ulam-Tsingou (FPUT) system and hard-particle gases \cite{VanBeijeren2012,Spohn2014,Popkov2016,Mendl2013,Narayan2002,Das2014}. 


One picture that has emerged from many studies is that, for systems with anomalous transport, the standard heat diffusion equation has to  be replaced by some fractional diffusion equation \cite{Lepri2016a,Bernardin2012,Lepri2009,Lepri2010,Delfini2010,Jara2015,Bernardin2016}. 
A particular model of anomalous transport where some rigorous results have been obtained is that of the harmonic chain whose Hamiltonian dynamics is supplemented by a stochastic part that keeps the conservation laws (volume, energy, momentum) intact --- we will refer this model as the harmonic chain momentum exchange (HCME) model. For the infinite HCME system, it was shown exactly that at equilibrium the energy current autocorrelation has a $\sim t^{-1/2}$ decay \cite{Basile2006}. 
It was also shown that, in contrast to Eq.~\eqref{heat-diff}, in infinite volume, the evolution of a localized energy perturbation $e(y,t)$, is described by a non-local fractional diffusion equation  
$\partial_t e(y,t) = -\bar{\kappa}(-\Delta)^{3/4} e(y,t)$,
where  $\bar{\kappa}$ is some constant which depends on microscopic parameters \cite{Jara2015}. The fractional Laplacian operator 
$(-\Delta)^{3/4}$ in the infinite space is defined by its Fourier spectrum: $|q|^{3/2}$ which  for the normal Laplacian operator $-\Delta \equiv -\partial^2_y$ is $q^2$. 

While most of the studies in HCME model consider evolution in infinite systems, it is also of interest to study transport across a finite system connected to two reservoirs of different temperatures at its two ends.
For diffusive systems in this set-up, the heat equation continues to describe both non-equilibrium steady state (NESS) and time-dependent properties. However, for anomalous transport, it is a priori not clear how to write a corresponding evolution equation in a finite domain. Since we expect this evolution to be governed by a fractional Laplacian which is a non-local operator, it is difficult to guess its representation in a finite system from its representation in the infinite system. Note that in the finite system one has to include the effects of the boundary conditions which are important as the operator itself is non-local. 
Hence, extending its definition 
to a finite domain is a non-trivial  problem. Several studies have addressed this problem of obtaining and studying fractional diffusion description in finite domain \cite{Viswanathan2000,Zoia2007, PrivateCommunication2015}. 

In this paper we study heat transfer across the HCME model connected to two reservoirs at its two ends. It has been observed and proved that in this model, heat current scaling with system size is anomalous and the steady state temperature profile is inherently non-linear \cite{Lepri2016a,Lepri2009,Cividini2017}. In the present work we provide a fractional equation description of the 
anomalous heat transfer both in the stationary as well as in the non-stationary state. Using this fractional description we derive new results related to evolution of temperature profile, equilibrium current fluctuations and to two point correlations in NESS. Below we summarise the main results of our work along with the plan of the paper:
\begin{itemize}
\item  In Sec.~\eqref{sec:Model} we first review  previous studies on the HCME model. These studies show that the macroscopic time evolution of two-point correlations is described by a set of coupled local linear PDEs \cite{Lepri2009,Lepri2010}. Starting from these PDEs, it can be shown that they naturally give rise to an evolution equation for the temperature profile $T(y, \tau)$ 
\[
\partial_\tau T(y,\tau)=-\bar \kappa~\mathbb L T(y,\tau),
\]
governed by a fractional Laplacian $\mathbb L$ defined in a finite domain, where $\tau$ is a scaled time (see later).  The operator $\mathbb L$ is defined in the domain $0 \leq y \leq 1$ through it's action \[\mathbb L \phi_n(y) = \lambda_n^{3/4} \phi_n(y)\]  on the complete Neumann basis $\iprod {y}{\phi_n}=\phi_n(y)=\sqrt{2}\cos(n \pi y)$ for $n \ge 1$ and $\iprod{y}{\phi_0}=\phi_0(y)=1$ with $\lambda_n=(n\pi)^2$.
Using this representation, we show that one can recover the exact results \cite{Lepri2009}  for the steady state temperature and current profiles in the HCME.

\item Next in Sec.~\eqref{sec:timeevolution} we discuss the time evolution of the temperature profile, starting from an arbitrary initial profile, to the long-time NESS profile. In order to solve the fractional diffusion equation with Dirichlet boundary conditions for arbitrary time we are required to find the eigenvalues and eigenvectors of the fractional operator  $\mathbb L$ with Dirichlet boundary conditions. We describe an efficient procedure to compute this eigensystem.  We also provide a detailed discussion of some  properties of the eigensystem that distinguish them from the eigensystem of the normal Laplacian operator with the same boundary conditions. 

\item Inspired by the fluctuating equations for energy evolution in diffusive systems\cite{Bernardin2012a,Spohn2012}, in Sec.~\eqref{sec:Noise} we extend the definition of the fractional equation to include fluctuations and noise in equilibrium such that fluctuation-dissipation relation holds locally. 

\item Using the fluctuating fractional equation description, in Sec.~\eqref{sec:GreenKubo}, we first verify the validity of the equilibrium Green-Kubo relation in finite systems where we encounter some interesting mathematical identities that we establish numerically. This motivates and enables us to study the long-range correlations in NESS in Sec.~\eqref{sec:Longrange}, where we propose a conjecture on the relation between these correlations and the Green's function of the operator $\mb{L}$.

\item Finally, in Sec.~\eqref{sec:conc} we conclude our paper.
\end{itemize}

	\section{Definition of Model and survey of earlier results}
	\label{sec:Model}
	
	
	We consider the so-called harmonic chain momentum exchange model (HCME), which 
	considers an added stochastic component in the usual Hamiltonian dynamics of a harmonic chain. The stochastic part is such that it preserves volume, momentum and energy conservation but the other conserved variables of the harmonic chain are no longer conserved. Thus the stochastic model restores ergodicity while preserving the important conservation laws. Here we are interested in the open system where the system is driven by two Langevin-type heat baths.  Specifically we consider 
	a  system  consisting of $N$ particles and  attached to two heat baths. The Hamiltonian plus heat bath part of the dynamics is described by the following equations 
	\eqa{
		\dot{q}_i =& ~p_i~,~~\dot{p}_i = \omega^2(q_{i+1}-2q_{i} + q_{i-1}),  ~~ 1<i<N, \nn \\
		\dot{p}_1=& ~ \omega^2(q_{2}-2q_{1})  -\lambda p_1 + \sqrt{2\lambda T_L} \eta_L,      \label{eq:eom} \\
		\dot{p}_N=& ~ \omega^2(q_{N-1}-2q_{N}) -\lambda p_N + \sqrt{2\lambda T_R} \eta_R ~,\nn
	} %
	where $\{q_i,p_i\}$, $i=1,2\ldots,N$, are the positions and momenta of the particles, $T_L, T_R$ are the temperatures of the left and the right Langevin baths and $\eta_L,\eta_R$ are Gaussian white noise terms.  Additionally there is a stochastic noise, such that the momenta of nearest neighbour particles are exchanged ({\emph{i.e} $p_{i+1} \leftrightarrow p_{i}$) at a rate $\gamma$.  
		For this model the two point correlation functions satisfy a closed set of equations. 
		
		Following \cite{Lepri2010} let us denote the possible correlation matrices by   ${\bf U}_{i,j}=\langle q_i q_j \rangle$,${\bf V}_{i,j}=\langle p_i p_j \rangle$, and ${\bf Z}_{i,j}=\langle q_i p_j \rangle$. 
		One can show that the time evolution of these correlation functions is given by linear equations involving only these set of correlations and source terms arising from the boundary driving \cite{Lepri2010}. 
 Let us also define the correlation ${\bf z}^+_{i,j}=\left({\bf Z}_{i,j}-{\bf Z}_{i-1,j}+{\bf Z}_{j,i}-{\bf Z}_{j-1,i} \right)/2$.
The most interesting physical observables involve the correlations $T_i={\bf V}_{ii}$, which can be taken as the definition of local temperature and the energy current $J=\omega^2 {\bf z}^+_{i,i+1} +(\gamma/2)~({\bf V}_{i+1,i+1}-{\bf V}_{i,i})$.  
		In the $N \to \infty$ limit,  one observes that the fields $T_i$ and ${\bf z}^+_{i,j}$ have the scaling forms 
 $T_i(t) = T(i/N,t/N^{3/2})$ and ${\bf z}^+_{i,j} = \frac{1}{\sqrt{N}} C(|i-j|)/{N}^{1/2},(i+j)/(2N),t/N^{3/2})$.  In terms of the following scaling variables $x=|i-j|/{N}^{1/2}, y=(i+j)/(2N), \tau=t/N^{3/2}$, it has been shown  in \cite{Lepri2010} that the fields $T(y,\tau)$ and $C(x,y,\tau)$ satisfy the following coupled set of PDEs:  
		\eqa{
			\gamma^2 \p_x^4 C(x,y,\tau) &= \omega^2 \p_y^2 C(x,y,\tau), \nn \\
			\p_y T(y,\tau) &= -2 \gamma \p_x C(x,y,\tau)|_{x\to0}, \label{PDE}\\
			\p_\tau T(y,\tau) &=  \omega^2 \p_y C(x,y,\tau)|_{x\to0}, \nn
		}
with boundary conditions $C(x,0,\tau)=C(x,1,\tau)=0,C(\infty,y,\tau)=0, \partial_x^3C(0,y,\tau)=0$ and $T(0,\tau)=T_L$ and $T(1,\tau)=T_R$ 
		where, the domain of variables are $x \in [0,\infty)$ and $y \in [0,1]$ [note that in \cite{Lepri2010} $y \in (-1,1)$]. 
To study the time-evolution of the fields $C(x,y,\tau)$ and $T(y,\tau)$, one has to subtract the steady state solutions $C_{ss}(x,y)$ and $T_{ss}(y)$ of the above equations (whose explicit forms are given in \cite{Lepri2009}).  
The boundary conditions suggest that one  expand the difference fields
using the complete Dirichlet basis $\iprod {y} {\alpha_n} =\alpha_n(y) = \sqrt{2}\sin(n \pi y)$ for $n\ge 1$
		\eqa{
			C(x,y,\tau) -C_{ss}(x,y)= \sum_{n=1}^\infty \hat{C}_n(x,\tau) \alpha_n(y) \label{Cexp},\\
			 T(y,\tau) -T_{ss}(y)= f(y,\tau)=\sum_{n=1}^\infty \hat{f}_n(\tau) \alpha_n(y)~. \label{Texp} 
		}
Following \cite{Lepri2010} one then gets (see \ref{app:matrixrep}) the following matrix equation for the evolution of the components $f_n$: 
	\eqa{
		\dot{\hat{f}}_n =&-\bar{\kappa}	\sum_{k=1}^\infty  \mb{L}_{nk} \hat{f}_k,~~~~n=1,2,\ldots,\infty,~ \\
		~\text{where}, ~ &\mb{L}_{nk}=  \left[\mc{T} \Lambda^{3/4} \mc{T^\dagger} \right]_{nk}, \label{Lnk}
	}
	with $\mc{T}_{nl}=\iprod{\alpha_n}{\phi_l}=\int_0^1 dy \alpha_n(y) \phi_l(y)$, where $\phi_m(y) = \sqrt{2}\cos(m \pi y)$ for $m > 0$, $\phi_0(y) = 1$ and   $\Lambda_{m l} = \lambda_{m}\delta_{m l}$  is a diagonal matrix with $\lambda_n=(n\pi)^2$. The constant $\bar{\kappa} = {\omega^{3/2}}/{(2\sqrt{2 \gamma})}$. Therefore, the function $f(x,\tau)$ with homogeneous boundaries $f(0,\tau) = f(1,\tau) = 0$ satisfies, \eqa{		\p_\tau f(y,\tau) = -\bar{\kappa} \mb{L} f(y,\tau)\label{fde1} .} 
	From Eq. \eqref{Lnk}, one notices that $\mb L_{nk}$ can be written as 
	\[\mb{L}_{nk}=\bra{\alpha_n}\mb{L} \ket{\alpha_k}=\bra{\alpha_n}\left[\sum_{m=0}^\infty \lambda_m^{3/4} \ket{\phi_m}\bra{\phi_m}\right] \ket{\alpha_k},~~\forall~~n,k= 1,2,\ldots, \infty\]
	which allows us to identify the action of the  operator $\mb L$ acting on the set of basis functions $\phi_m$ (which satisfy Neumann boundary conditions):
	\eqa{\mb L \ket{\phi_m}= \lambda_m^{3/4} \ket{\phi_m}~. \label{L-in-phi}} 
	 It is important to notice that the above representation of the operator $\mathbb L$ is not the ``spectral fractional Laplacian with Dirichlet boundary conditions" which would consist to replace $\phi_n$ by $\alpha_n$ in \eqref{L-in-phi}.  This definition has been mentioned in \cite{Lepri2016a} and a more mathematically rigorous derivation has been obtained \cite{PrivateCommunication2015}.
         The above results imply that the temperature field $T(y,\tau)$ evolves according to the fractional equation 
\eqa{ 
\p_\tau T(y,\tau) = -\bar{\kappa} \mb{L} T(y,\tau)= -\mb{L}_{\bar{\kappa}} T(y,\tau)~, \label{fde}
}
where we have defined $\mb{L}_{\bar{\kappa}}=\bar{\kappa} \mb{L}$ and the steady state is required to satisfy the condition  $\mb{L}_{\bar{\kappa}} T_{ss}(y) = 0$. 
To describe the evolution of the temperature profile, one is specifically interested in finding the eigenvectors of the operator $\mb L$ which satisfy Dirichlet boundary conditions. This can be obtained by diagonalizing the infinite-dimensional matrix in Eq.~(\ref{Lnk}). Let the eigenvector components of this matrix be denoted by $\psi^{(m)}_n$, corresponding to   eigenvalue  $\mu_n$, so that $\sum_k \mb L_{mk} \psi^{(k)}_n= \mu_n \psi^{(m)}_n$.
Then the eigenvector in the position basis is given by $\psi_n(y)=\sum_m \psi_n^{(m)} \alpha_m(y)$.
 In Sec.~\ref{sec:timeevolution} we provide an alternate and more efficient method of computing eigenvalues and eigenvectors. This method involves finding roots of a transcendental equation and avoids diagonalization of  infinite dimensional matrices. We also discuss various properties of the spectrum there. 
		We now describe several results that follow for the steady state and the time evolution towards it.
		
		\subsection{Steady state results}
		
Let us write the steady state temperature in the form
\be
T_{ss}(y) = \overline{T} + \delta T ~\Theta(y),
\ee
where $\overline{T} = {(T_L+T_R)}/{2}$,  $\delta T = T_L -T_R$ and the function $\theta(y)$ satisfies the  boundary conditions, $\Theta(0) = 1/2$, $\Theta(1) = -1/2$~. Then  expanding  $\Theta(y) = \sum_n \hat{\Theta}_n \phi_n(y) $, the stationarity condition 	$\flap \Theta =0 $ along with Eq.~\eqref{L-in-phi} gives
\eqa{
  \sum_n \lambda_n^{3/4} \hat{\Theta}_n \phi_n=0. \label{stateq}
}
Now we note the identities (see \ref{app:cossumidentity}), which have to be understood in a distributional sense:
\eqa{ \sum_{n\;  odd} \phi_n(y) = 0,~
  \sum_{n\;  even} \phi_n(y) = -{1}/{\sqrt{2}} \label{cosid}.} Using these and the boundary conditions  $\Theta(0)=-\Theta(1) = \frac{1}{2}$ we finally get
\eqa{
  \Theta(y) &= \sum_{n \; odd} \frac{c}{\lambda_m^{3/4}} \phi_m(y), \nn \\
	{\rm with}~~c&=  \frac{\pi^{3/2}}{[\sqrt{8} -1] \zeta(3/2)},
}
where $\zeta(s)$ is the Riemann-Zeta function. The temperature profile matches with the one presented in \cite{Lepri2016a,Lepri2009, Cividini2017}:
\eqa{
  T_{ss}(y) =& \overline{T} + \delta T \frac{ \pi^{3/2}}{[\sqrt{8}-1  ] \zeta(3/2)} \sum_{n \; odd} \frac{\phi_n(y)}{\lambda_n^{3/4}}.
  \label{eq:steadystate}}
A comparison of the above equation with the microscopic simulation of the system Eq.~\eqref{eq:eom} in Fig.~\eqref{fig:tempprofile} shows a very good agreement. The systematic differences are due to finite size effects, as was already noted in \cite{Lepri2009}.
		We next consider the steady state current. First, we observe that the fractional Laplacian $\flap$ can be expressed in the form of a divergence, namely in the form $\flap=\bar{\kappa} \p_y \mb{A}$ where the operator $\mb{A}$ is defined through the following action on Neumann basis vectors
		\eqa{
			\mb{A} \phi_n(y) = \lambda_n^{1/4} \alpha_n(y)~.\label{eq:current}
		}  
	We then see that \eqref{fde} is in the form of a continuity equation $\p_\tau T(y,\tau)= -\p_y j(y,\tau)$ with the non-local energy current defined as $j(y,\tau)=\bar{\kappa} \mb{A} T(y,\tau)$.
		Using this definition of the current and the steady state temperature profile in \eqref{eq:steadystate} we immediately get the steady state current as
		\eqa{
			\frac{j}{\delta T}= \frac{\bar{\kappa} \mb{A} T_{ss}(y)}{\delta T}  = \frac{\bar{\kappa} c}{2 \sqrt{2}}~, \label{eq:sscurrent}
		}
		where we used the identity $\sum_{n \in {\rm odd}} \alpha_n(y)/\lambda_n^{1/2}=1/(2 \sqrt{2})$ \eqref{app:cossumidentity}.  Note that this gives us the scaled current while the actual current is given by $J=j/\sqrt{N}$, in agreement with results obtained in \cite{Lepri2009}.

		\begin{figure}
			\centering
			\includegraphics[width=0.7\linewidth]{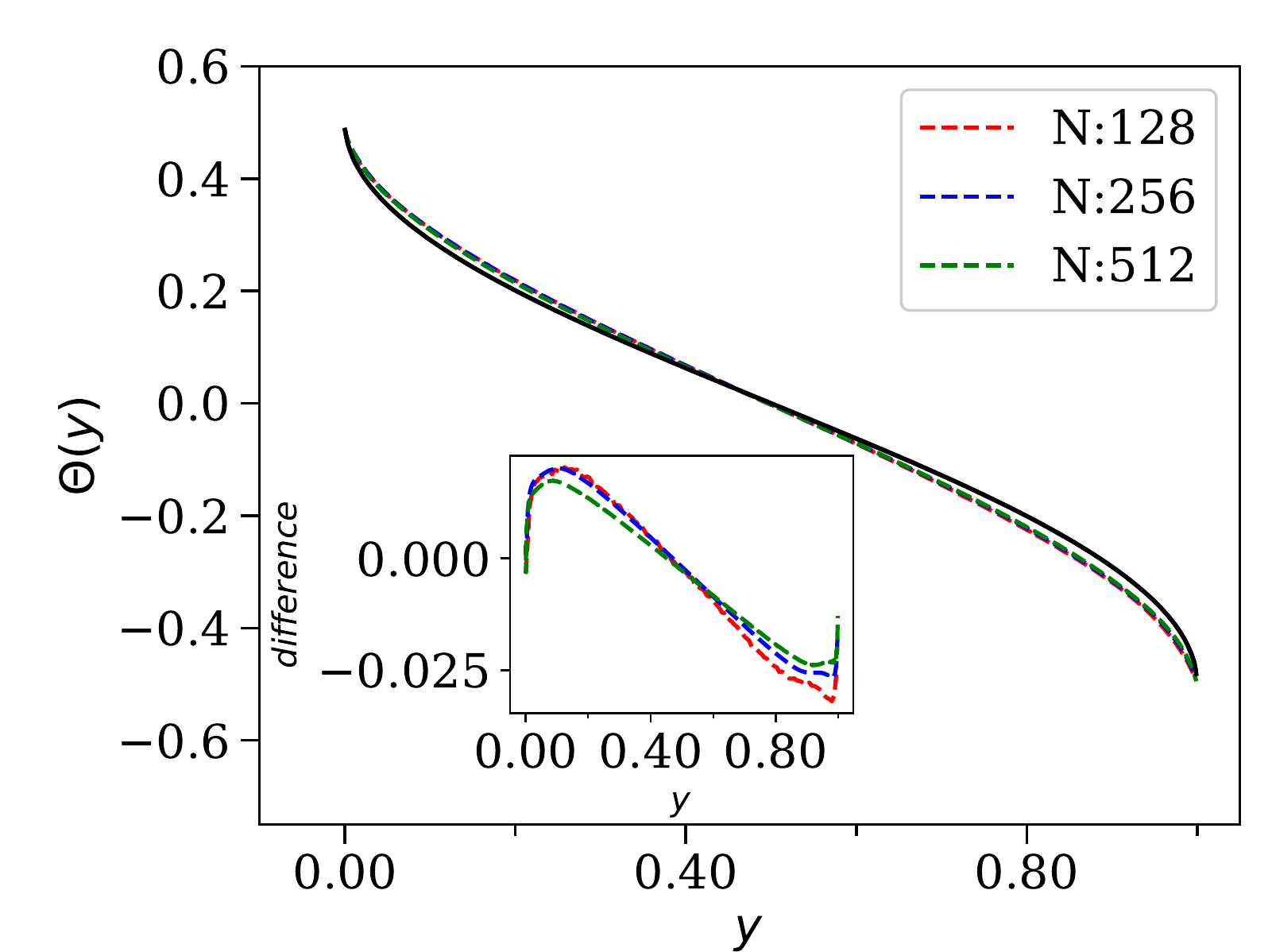}
			\caption{Temperature profile from Eq.~\eqref{eq:steadystate}(solid black line)  compared with direct numerical simulations of microscopic system for system sizes $N = 128,256,512$. In the inset the difference between Eq.~\eqref{eq:steadystate} and numerical simulations is plotted for various system size. }
			\label{fig:tempprofile}
		\end{figure}

		
		\section{Time evolution of temperature profile}\label{sec:timeevolution}


		The fractional Laplacian equation \eqref{fde}  allow us to study the time evolution of the temperature profile, starting from  given initial and boundary conditions, and the eventual approach to the steady state at large times. Here we address the problem of  describing  the system's time evolution. As before, the temperature profile at any time $\tau$ in the form $ T(y,\tau)=T_{ss}(y)+\tdir(y,\tau) $, where again $\tdir(y,\tau)$  satisfies Eq.~\eqref{fde} but with vanishing Dirichlet boundary conditions, $\tdir(0,\tau) = \tdir(1,\tau) = 0$.  
		Let $\{ \psi_n\}$ be the eigenvectors with corresponding eigenvalues $\mu_1 < \mu_2 < \mu_3 \dots$ of $\flapfinite$ , satisfying the equation		\eqa{
		\flapfinite  \psi_n(y) = \mu_n  \psi_n(y) 
		\label{eq:fraceval}
		}
		and boundary conditions $\psi_n(0)=\psi_n(1)=0$. 
		It can be shown that the  operator $\flapfinite$ has a non-degenerate and positive spectrum (see below).  We can then immediately write 
		the solution for  $\tdir(y,\tau)$  as
		\eqa{\tdir(y,\tau) = \sum_{n=1} \hat{\tdir}_n(0) e^{-\bar{\kappa}\mu_n \tau } \psi_n(y), \label{eq:timeevolution} \\ 
			~ {\rm where} ~~ \hat{\tdir}_n(0) = \int_{0}^{1} dy ~\tdir(y,0) \psi_n(y), \nn
		}
		are ``fractional-Fourier coefficients" for the initial field $\tdir(y,0)$.  
		In the first section we outlined the procedure followed in \cite{Lepri2010} to find the Dirichlet eigenfunctions  expanding the  eigenfunctions $\psi_n$ in the orthogonal basis of $\{ \alpha_l \}_{l\ge 1}$ as  $\psi_n(y) = \sum_{l\ge 1} \xi_{nl} \alpha_l(y)$. 
		We show here that much simplification and better accuracy is achieved if one expands $\psi_n$ directly in the Neumann basis $\{\phi_m\}_{m\ge 0}$. 
		\eqa{
			{\psi}(y) = \sum_m \hat{\mzeta}_m {\phi_m}(y).
		}
		From Eq.~\eqref{eq:fraceval}, and using the definition of $\flapfinite$ in Eq.~\eqref{L-in-phi},  we have
		\eqa{
			\sum_{m\ge 0} (\mu - \lambda_m^{3/4}) \hat{\mzeta}_m {\phi_m}(y) = 0.
			\label{eigeqn}
		} 
		There are two sets of solution for this equation. The first set is given by 
		\eqa{
			\hat{\mzeta}_0= - \frac{b}{\sqrt{2}\mu},~ 
			\hat{\mzeta}_{2k} = \frac{b}{\lambda_{2k}^{3/4} - \mu},~~ k \ge 1,~~ 
			\hat{\mzeta}_{2k+1} = 0,~~ k \ge 0~, \label{SolS1}
		}
		where we have made use of the identity $\sum_{m=1}^\infty \phi_{2m}(x)=-1/\sqrt{2}$.
		The second solution set is given by
		\eqa{
			\hat{\mzeta}_{2k+1} = \frac{b}{\lambda_{2k+1}^{3/4} - \mu},~~ k \ge 0, ~~~
			\hat{\mzeta}_{2k} = 0,~~ k \ge 0,\label{SolS2}
		}
		where we have used the identity $\sum_{m=0}^\infty \phi_{2 m+1}(y)=0$ \eqref{app:cossumidentity}.
		So far,  $b$ and $\mu$ are un-determined. We now use the Dirichlet boundary condition 
		$\psi(0) =  \hat{\mzeta}_0 + \sqrt{2} \sum_{k\ge1} \hat{\mzeta}_{2k} + \sum_{k\ge0} \hat{\mzeta}_{2k+1} = 0$.   From our first solution set Eq.~\eqref{SolS1} we then get the following equation satisfied by $\mu$
		\eqa{
			\sum_{k\ge1}  \frac{1}{\lambda_{2k}^{3/4} - \mu} = \frac{1}{2\mu}~.
			\label{eq:eval1}
		}
		Similarly, from the second solution set Eq.~\eqref{SolS2}, we get
		\eqa{
			\sum_{k\ge0}  \frac{1}{\lambda_{2k+1}^{3/4} - \mu} = 0~.
			\label{eq:eval2}
		}
		The solution of either of the above two equations gives us the required eigenvalue, while Eqs.~\eqref{SolS1}-\eqref{SolS2} provide us with the corresponding eigenfunction, with the constant $b$ fixed by normalization. 
		We label the first set of solutions by $\mu_{2n+1},\psi_{2n+1}$, $n\ge 0$  and the second set by $\mu_{2n},\psi_{2n+2}$, $n\ge 0$. 
		From the  structure of the eigenvalue equations  it is clear that the roots are ordered set of numbers such that $\lambda_{2n}^{3/4} < \mu_{2n+1} < \lambda_{2n+2}^{3/4}$ and $\lambda_{2n-1}^{3/4} < \mu_{2n} < \lambda_{2n+1}^{3/4}$. 
		Finally, introducing the notation, $\iprod{f}{g} = \int_0^1 dx' f(x')g(x')$, such that $\iprod{x}{\psi_n} =  \int_0^1 dx \delta(x-x')\psi_n(x') = \psi_n(x)$, the eigenvectors can now be written explicitly as
		\eqa{
			\ket{\psi_{2n+1}} &= D_{2n+1} \left( -\frac{1}{\sqrt{2}\mu_{2n+1}}\ket{\phi_0}  + \sum_{m\ge1} \frac{1}{\lambda_{2m}^{3/4} -\mu_{2n+1}} \ket{\phi_{2m}} \right), \\
			\ket{\psi_{2n+2}} &= D_{2n+2} \left( \sum_{m\ge0} \frac{1}{\lambda_{2m+1}^{3/4} -\mu_{2n+2}} \ket{\phi_{2m+1}} \right) ,
			\label{eveceta}
		}
		where $D_{n}$, found from the normalizing condition $\iprod{\psi_{n}}{\psi_{n}} =1$, is explicitly given as,
		\eqa{
			D_{2n+1} &= \left[\frac{1}{2\mu^2_{2n+1}} + \sum_{m\ge1} \frac{1}{(\lambda_{2m}^{3/4} -\mu_{2n+1})^2}\right]^{-1/2},\nn\\
			D_{2n+2} &= \left[\sum_{m\ge0} \frac{1}{(\lambda_{2m+1}^{3/4} -\mu_{2n+2})^2}\right]^{-1/2}~.
		} 
		Thus, as promised, we have managed to obtain a much efficient method for computing the Dirichlet spectrum of the fractional operator $\mb{L}$. The roots of the eigenvalue equations \eqref{eq:eval1}-\eqref{eq:eval2} are solved numerically using Newton-Raphson method scanning in between these intervals. 
		This procedure gives a fast and efficient way to compute the eigenvector while avoiding diagonalizing infinite dimensional matrices. For large $k$, we have, $\mu_k \approx \lambda_k^{3/4}$. 
		
This procedure can be generalized to a fractional operator defined through the equation
\eqa{
\flapfinite^{(\genpower)}  \phi_n(y) =  \lambda_n^{\genpower} \phi_n(y), \label{Lbeta}
}
for  arbitrary $\genpower$. For diffusive case ($\genpower = 1$) one can obtain exact results and recover  the expected result $\mu^{(\genpower=1)}_{k} = \pi^2 k^2$ and $\psi_k(y)=\alpha_k(y)$.

		\subsection{Properties of Dirichlet eigensystem of the fractional operator in bounded domain}

	\begin{figure}
			\centering
			\includegraphics[width=0.7\linewidth]{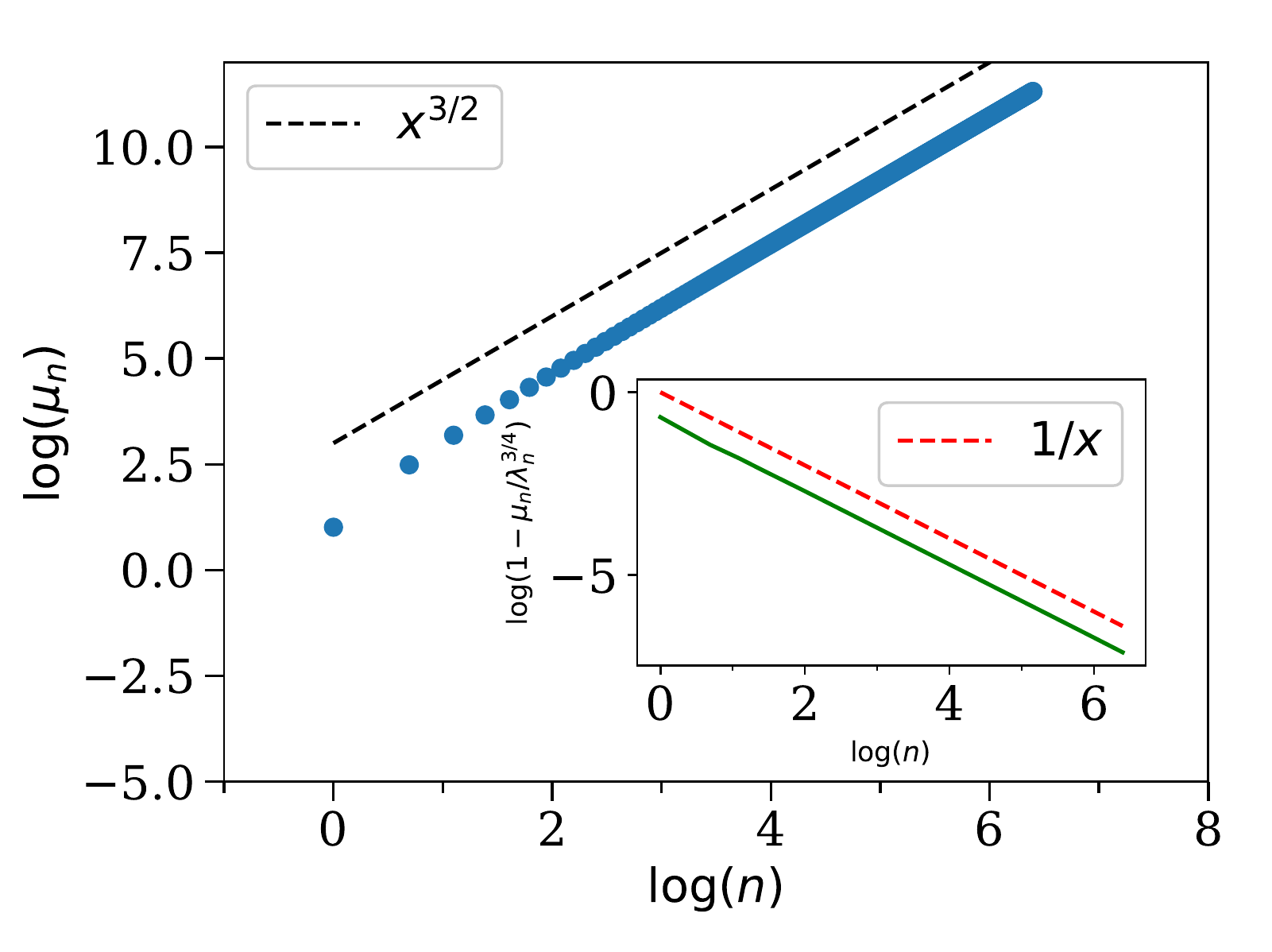}
			\caption{Eigenvalues computed from Eq.~\eqref{eq:eval1} and  \eqref{eq:eval2} plotted (blue dots) in log-log scale. Number of basis state used to approximate the function is $600$. For large $n$, $\mu_n \sim (n\pi)^{3/2}$, i.e. the Dirichlet and Neumann eigenvalues are same. For small $n$ there is a systematic difference between the two. A straight line of exponent $x^{3/2}$ (black dot) is plotted alongside. In the inset we plot, $\log(1-\mu_n/\lambda_n^{3/4})$ vs $\log(n)$, which characterizes the difference between the Dirichlet and Neumann boundary eigenvalues.  For large $n$ the value of this function goes to zero with slope $1$, suggesting he difference between the two decreases linearly with $n$. The red dashed line shows that it decays with an inverse power law of exponent $1$.  }
			\label{fig:eigenvals}
		\end{figure}

		\begin{figure}
			\centering
			\includegraphics[width=0.7\linewidth]{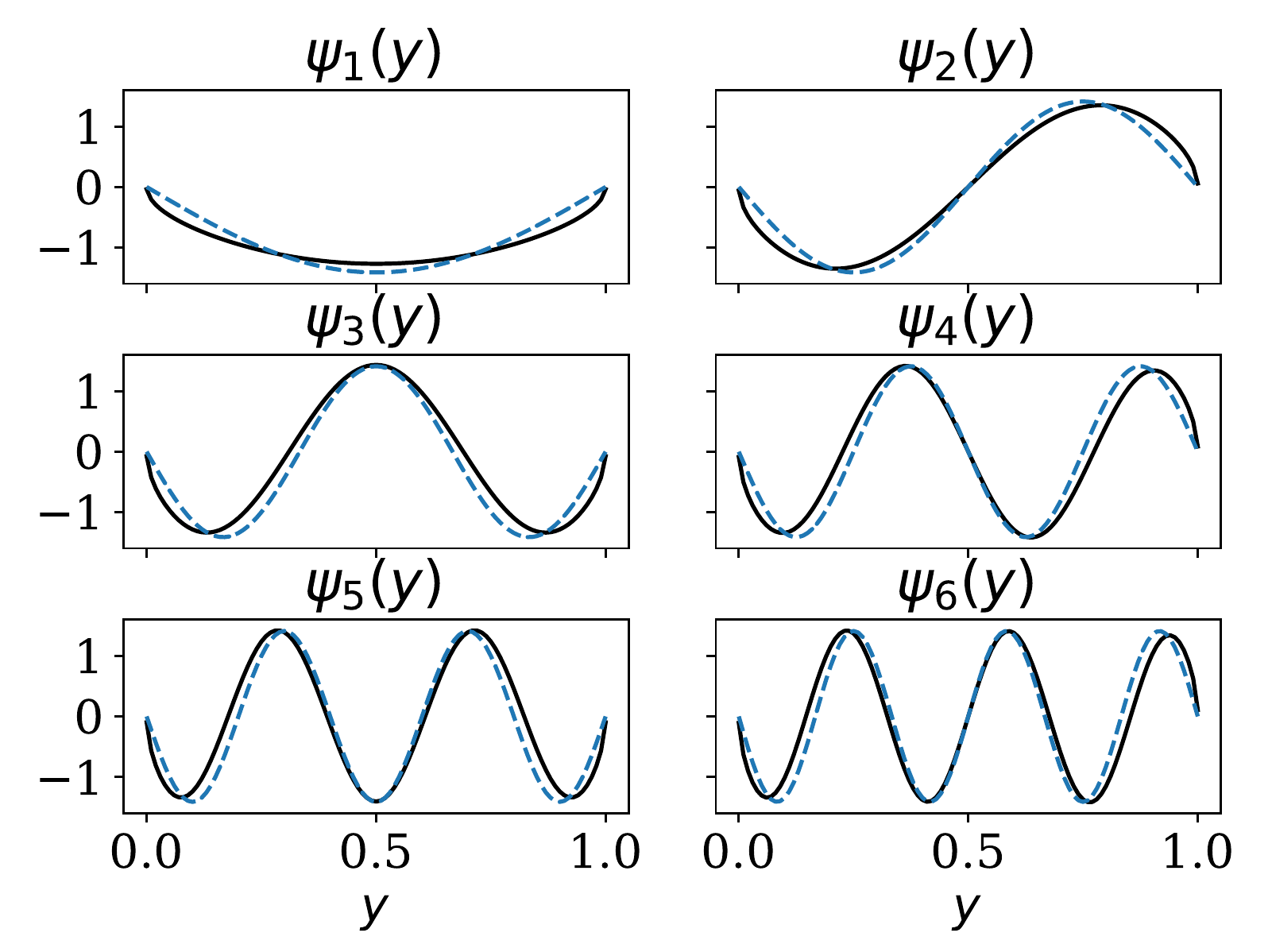}
			\caption{The first six eigenvectors of fractional operator, $\psi_n(y)$, (black) as compared to corresponding eigenfunctions of Laplacian i.e. $\sin$ functions (blue dotted). The eigenstates are different from corresponding $\sin$ functions near the boundaries even for large $n$.  These eigenfunctions are computed by summing over $600$ basis states.}
			\label{fig:eigenvectors}
		\end{figure}
	
	 The numerical values of the computed eigenvalues are plotted in Fig.~\eqref{fig:eigenvals} in log-log scale, where we find that for large $n$  $\mu_n \approx (n \pi)^{3/2}$, while for smaller values $n$, there is a systematic deviation from the scaling due to the fact we are now working in a bounded domain. The first three eigenvalues ($\mu_n$) are approximately $\mu_1 \approx 2.75,\mu_2 \approx 12.02, \mu_3 \approx 24.22$. The first eigenvalue we have $|\mu_1-\pi^{3/2}|/\pi^{3/2} \approx 0.5046$ (see inset in Fig.~\eqref{fig:eigenvals}). This eigenvalue spectrum is expected to be identical to that in \cite{Lepri2010}, upto a constant factor (see discussion in previous section). The first few numerically computed eigenvectors are shown in Fig.~\eqref{fig:eigenvectors}.  The eigenvectors are similar to $\sin$ functions but have divergent derivatives near the left and right boundaries. In order to compare it with corresponding $\sin$ functions, we plot in Fig.~\eqref{fig:eigenvectorsprop} the overlap of integral between $\psi_n(y)$ and  $\sqrt{2}\sin(n \pi y)$ defined as $I_n=1- \int_0^1 \psi_n(y)\sqrt{2}\sin(n \pi y) ~dy$. This increases and saturates to a particular value, suggesting that the wave functions are quite different from $\sin$ functions even for large $n$. Also the eigenfunctions show a non-analytic behavior at the boundaries, for example  near the left boundary one finds $\lim_{y\to 0^+}\psi_n(y) \sim \sqrt{y}$ (see Fig.~\ref{fig:eigenvectorsprop}b),  in contrast to $\sin$-functions for which  $\lim_{y\to 0^+}\sin(n \pi y) \sim y$. 
		\begin{figure}
			\centering
			\includegraphics[width=0.47\linewidth]{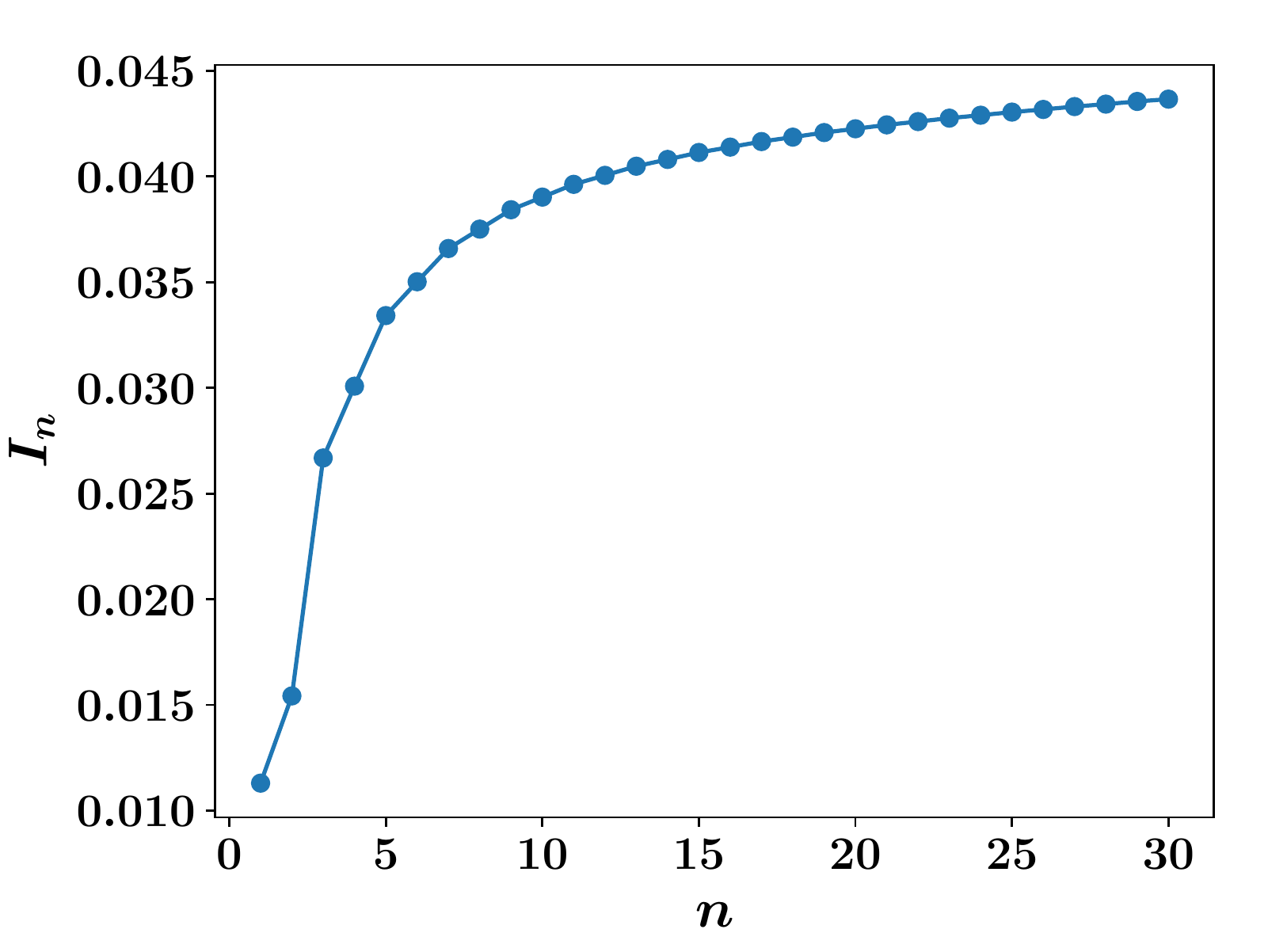}
			\includegraphics[width=0.47\linewidth]{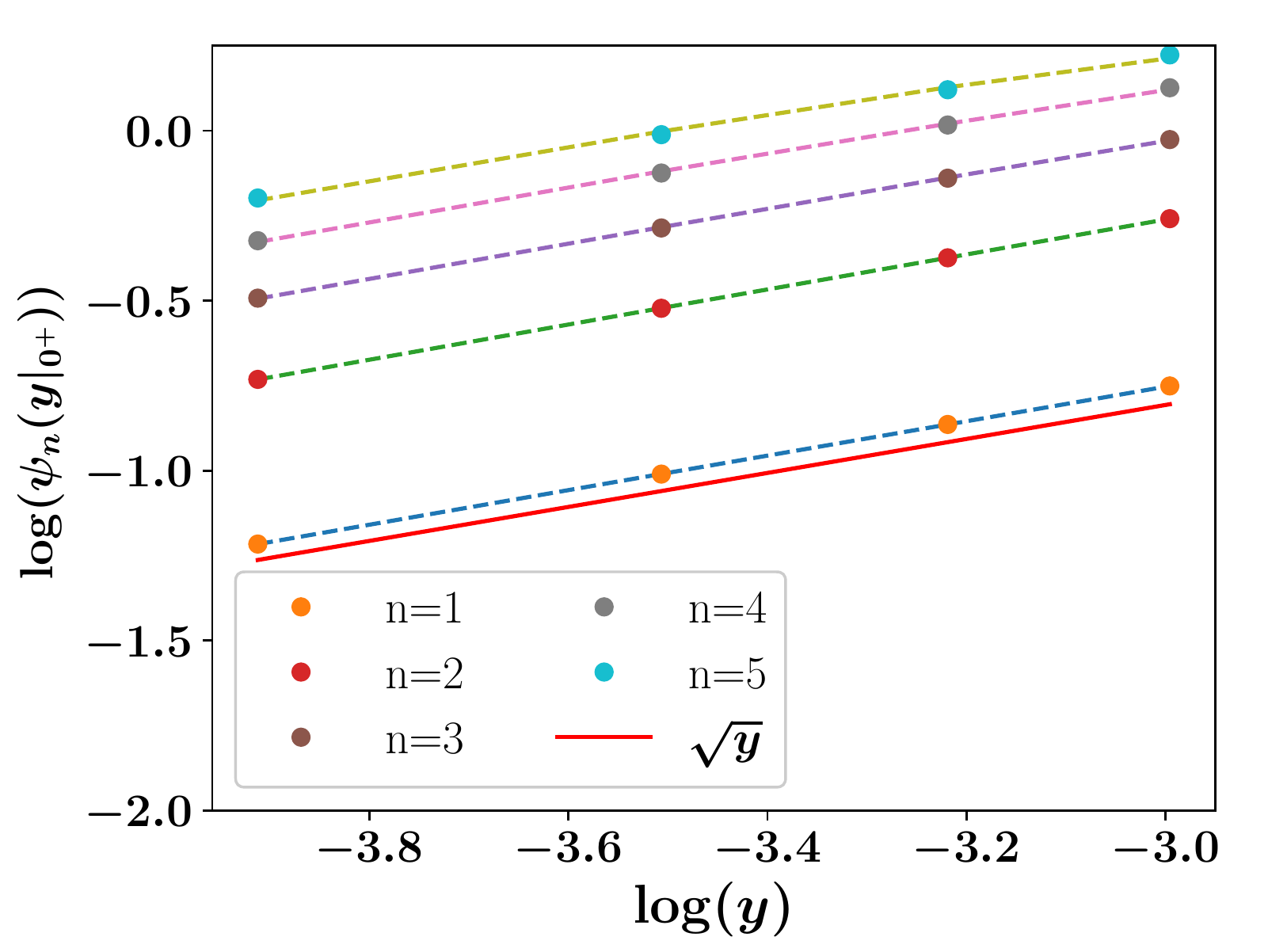}
			\caption{(a)(Left) To quantify the similarity between $\psi_n(y)$ and $\sqrt{2} \sin(n \pi y)$, we plot the overlap integral, $I_n=1- \int_0^1 \psi_n(y)\sqrt{2}\sin(n \pi y) ~dy$ . For  large $n$ this seems to be saturating to a finite value, suggesting that the eigenfunctions $\psi_n$ are quite different from $\sin$ functions. (b)(right) Scaling at boundaries of the eigenvectors in log-log scale shows that at the boundaries, the wave-function scales as $\sqrt{y}$. }
			\label{fig:eigenvectorsprop}
		\end{figure}

	The eigenspectrum of fractional operator in bounded domain has  been 
discussed earlier in the literature, using somewhat phenomenological approaches   \cite{Zoia2007,Buldyrev2001,Buldyrev2001a,Chen2004}. It is not clear if those approaches can be  related to that presented in this paper. 

		\subsection{Comparison of time evolution formula with numerical simulations of the HCME model}
		We now compare the prediction from Eq.~\eqref{eq:timeevolution}, with $\bar{\kappa} =1/(2\sqrt{2})$, with results from direct microscopic simulations, described by Eq.~\eqref{eq:eom} with the additional stochastic exchange  dynamics. Initially the system of size $N$ is prepared in a step initial condition, given by
\eqa{T_i &= T_L ,~ 1\le i <N/2,  \nn\\
&= T_R,~ N/2\le i\le N+1.
}
At large times it reaches a steady state described by Eq.~\eqref{eq:steadystate}.  At various intermediate times, we plot the function $\Theta(y,\tau) = T(y,\tau) - \overline{T} = T_{ss}(y)+f(y,\tau) - \overline{T}$, such that $\Theta(0,\tau) = 1/2 =-\Theta(1,\tau)$. 
In Fig.~\eqref{fig:timeevolution} we show the temperature profile at intermediate times from microscopic simulation with scaled space ($y=i/N$) and times ($\tau=t/N^{3/2}$) for various system sizes. We note that with increasing  system size, the data converges to the prediction from Eq.~\eqref{eq:timeevolution}. The difference between the numerical profiles and the predicted theoretical profile is shown in the inset.  As we increase the system size, this difference systematically decreases. We also demonstrate that using  standard Dirichlet $\sin$-functions, instead of the $\psi$-functions, leads to significant differences, especially near the boundaries.
		\begin{figure}
			\centering
			\includegraphics[width=0.45\linewidth]{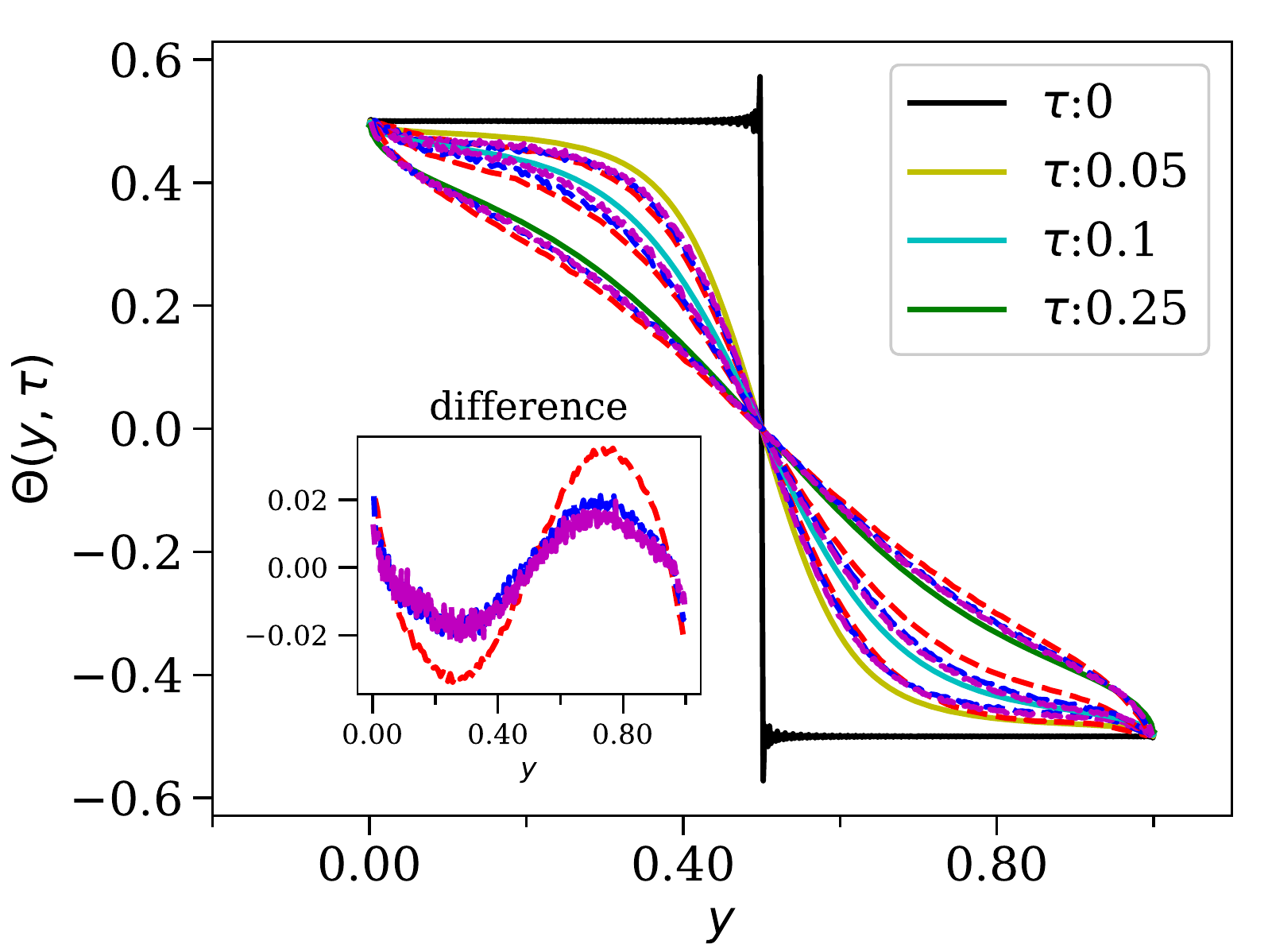}
			\includegraphics[width=0.45\linewidth]{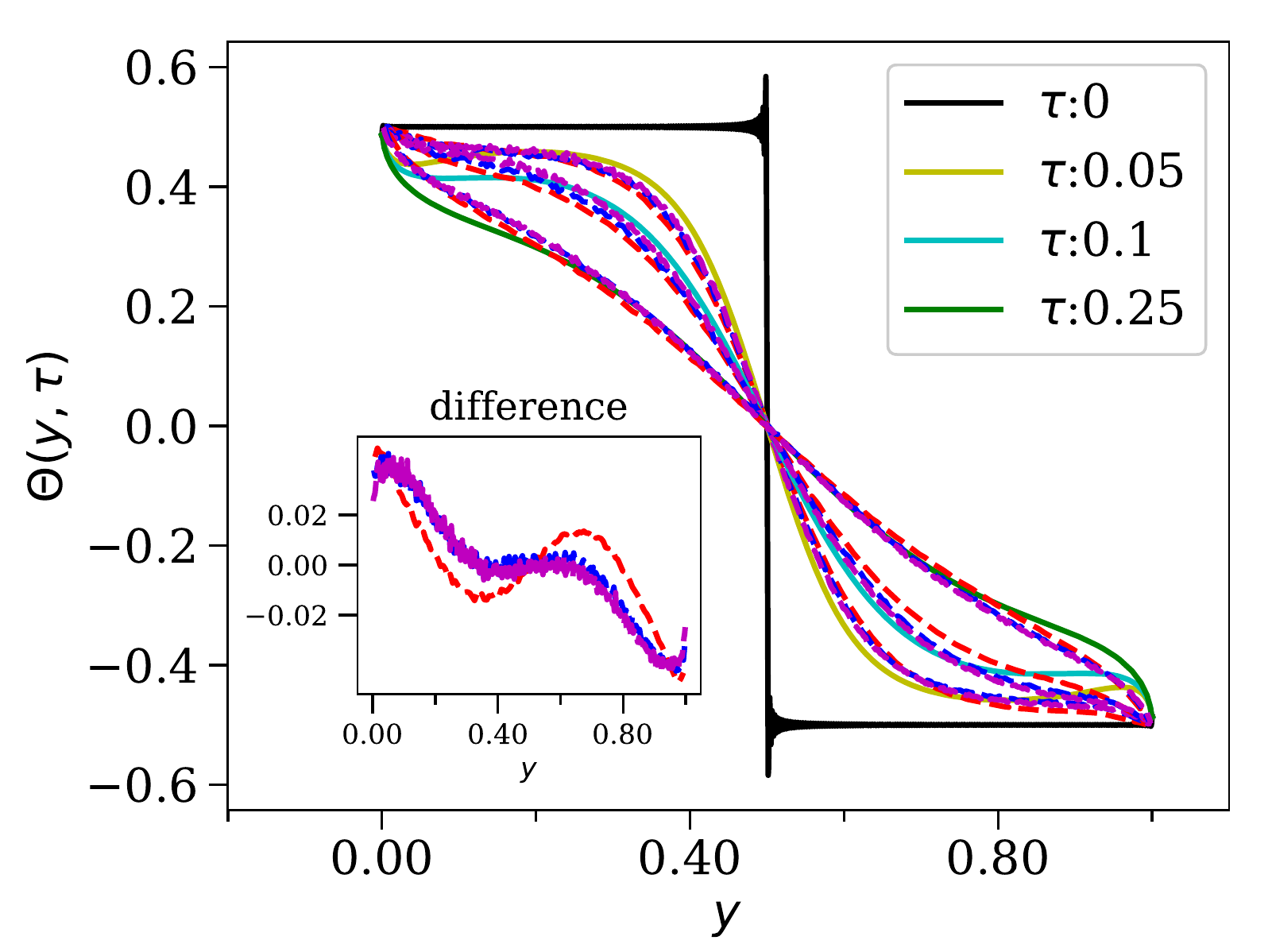}
			\caption{The time evolution of temperature starting from an initial step profile. The function $\Theta(y,\tau) = T(y,\tau) - \overline{T} = T_{ss}(y) + f(y,\tau) -\overline{T}$ is plotted and compared with  numerical simulations. In the left figure, dashed lines indicate  simulation results for the time-evolution, for system sizes  $N=128$ (red), $N=256$ (blue), $N=512$ (magenta). The solid lines at different scaled times ($\tau$) are generated from  Eq.~\eqref{eq:timeevolution} by summing over $600$ basis states. (right) The same, but now with the theoretical curves computed using the $\sin$-functions instead of the $\psi$-functions,  and eigenvalues $\lambda_n^{3/4}$ instead of $\mu_n$. We  notice that they do not match well with simulations, specially the deviations are prominent near the two boundary.}
			\label{fig:timeevolution}
		\end{figure}

		
		\section{Adding noise satisfying fluctuation dissipation to describe equilibrium fluctuations in finite system}
		\label{sec:Noise}
		

		In \cite{Bernardin2012a}, the harmonic chain with random momentum flips (HCMF model) was studied.  In the HCMF, the stochastic dynamics flips the momentum of the particle and is embedded in the Hamiltonian dynamics such that the macroscopic dynamics is diffusive. It was shown that the equilibrium energy fluctuations $e(x,t)=E(x,t)-\la E(x,t)\ra$, where  $E(x,t)$ is the local energy of the system at time $t$, satisfies the noisy diffusion equation $\p_t e(x,t) = \p_{x}^2e(x,t) + \p_x( D T(x,t) \eta(x,t))$,  with $\eta$ a space-time mean zero white noise. As a note of caution, from here onwards, we use a different notation than that of the previous sections with $\{x,y\} \in [0,1]$, and $t$ is, in general, referred to a scaled time. 
The aim of this section is to establish a fractional fluctuating equation for the HCME model, which has anomalous diffusion properties. Using this equation, we establish a Green-Kubo formula relating the equilibrium current fluctuations to the non-equilibrium current. Next, we discuss the long-range correlations and conjecture a form for the long-range correlations of energy and test it using simulations.
		
		The generalized equation, which we hypothesize in equilibrium at temperature $T$ is
\eqa{
			\p_t \ket{e_t} &= -\flap \ket{e_t} + \sqrt{ 2 \bar{\kappa}} \nabla (B T \ket{\eta_t}),
\label{eq:fluctuatingeq}
}
where $\eta(x,t)$ is a white Gaussian noise with $\blang \eta(x)\brang=0$, $\blang \eta(x,t) \eta(y,t') \brang = \delta(x-y)\delta(t-t')$ and $ \flap$ is the fractional Laplacian as defined in Eq.~\eqref{fde}.  The explicit form for the operator $BB^\dagger$ is established through the requirement that energy fluctuations must  respect the fluctuation dissipation (FD) in equilibrium. We define the Green function satisfying
		\eqa{\p_t G_t =  -\flap G_t, ~   \langle x| G_0 |x' \rangle= G^{xx'}_0=\delta(x-x'), \label{eq:greenfunction}}    
		with Dirichlet boundary conditions in $x \in [0,1]$. This can then be easily expressed in terms of the basis states $\{\psi_n\}_{n\ge 1}$ as  $ G^{xx'}_t = \sum_{n=1} \psi_n(x)\psi_n(x') e^{-\bar{\kappa} \mu_n t} $.
		The long-time solution to Eq.~\eqref{eq:fluctuatingeq} is then given by
		\eqa{e(x,t)=& \sqrt{ 2 \bar{\kappa}}  \int_{-\infty}^t ds  \braket{x}{G_{t-s}}{\nabla (BT\eta_s)},\nn\\
			=&- \sqrt{ 2 \bar{\kappa}} \int_{-\infty}^t ds  \braket{x}{\nabla G_{t-s}}{(BT\eta_s)}.}
		The equal time correlation function in equilibrium defined as $C_{\rm eq}(x,y) = \blang e(x,t) e(y,t) \brang $ then is given as
		\eqa{C_{\rm eq}(x,y) =& 2 \kappa \int_{-\infty}^t ds ~\int_{-\infty}^t ds' \blang \braket{x}{\nabla G_{t-s}}{BT\eta_s}  \braket{\eta_{s'} T B^\dagger }{\nabla G_{t-s'}}{y} \brang ,\nn  \\
			=& ~  2\bar{\kappa} T^2  \int_{-\infty}^t ds ~ \braket{x}{\nabla G_{t-s}BB^{\dagger}\nabla G_{t-s} } {y},\\
			=&  - 2\bar{\kappa} T^2  \int_{-\infty}^t ds  \braket{x}{ G_{t-s}\nabla BB^{\dagger}\nabla G_{t-s} } {y} ,\nn}
		where the statistical average is used and we  integrate out the space time white noise to give
		$ \blang \iprod{x'}{ B\eta_s}\iprod{\eta_{s'} B^\dagger}{y'}   \brang =  \blang {B\eta(s)} B\eta (s') \brang = \delta(s-s') BB^\dagger(x',y')$ followed by an integration by parts.  Here, the big angles, $\blang \dots \brang$ denote average over space-time white noise profiles whereas $\langle..|$ and $|..\rangle$ denote the bra-ket notation, \emph{e.g.} $\langle x|e \rangle =e(x,t)$. If we identify $ -\bar{\kappa} \nabla BB^{\dagger}\nabla  = \flap$, and using Eq.~\eqref{eq:greenfunction}  we recover FD relation in equilibrium 
\eqa{
C_{\rm eq}(x,y)  =&  ~  T^2  \int_{-\infty}^t ds ~\left( \braket{x}{ G_{t-s}\flap G_{t-s} }{y} +  \braket{y}{ G_{t-s}\flap G_{t-s} }{x} \right),\nn \\
			=&~ T^2  \int_{-\infty}^t ds ~\p_s   \braket{ x}{G_{t-s}G_{t-s} }{y} 
			= ~T^2 \delta(x-y),
}
		where we used the fact that changing $x \leftrightarrow y$ would not change the correlation function due to time reversal symmetry of the microscopic dynamics. The operator $BB^\dagger$  can consistently be defined on a function $g(x)$, expanded in $\{\alpha_n\}_{n\ge 1}$ basis,
		as $g(x) = \sum_n \hat{g}_n \alpha_n(x)$. Again, using the definition of $\flap=-\bar{\kappa} \nabla BB^{\dagger}\nabla$ in Eq.~\eqref{L-in-phi}, we define the symmetric operator $BB^\dagger$ as,
		\eqa{ \int_0^1 dx' BB^\dagger(x,x') g(x')  = \sum_{n=1}^{\infty} \frac{1}{(\lambda_n)^{1/4}} \hat{g}_n \alpha_n(x). \label{eq:BBdagger}} 
		Note that we do not assume anything about the form of the operator $B$, which would be important if we were to study non-equilibrium phenomena where temperature is not constant in space.
		
		The connection between the $\mathbb{L}$ operator with the $BB^\dagger$  allows one to identify the  
		 current (through continuity equation) as 
		\[
		j(x,t)=  -\bar{\kappa} \int_0^1 dx'~ BB^\dagger(x,x') \partial_{x'}e(x',t).
		\]
		Note that the above equation is a linear response relation, but in contrast to the diffusive case, this relation is non-local. Such non-local linear response relation have  recently been reported in \cite{Cividini2017}, where an alternate series representation of the kernel $BB^\dagger(x,x')$ has been provided for HCME model with general boundary conditions. 
		In  \ref{app:BBdaggergen}, we show that the spectral representation in Eq.~\eqref{eq:BBdagger} is completely consistent with the series representation in \cite{Cividini2017} for fixed boundary condition.

		\subsection{Spatio-temporal equilibrium energy correlations}
		We compute the two time spatio-temporal energy correlations in equilibrium defined as  $C_{\rm eq}(x,t,y,t') = \blang e(x,t) e(y,t') \brang$ and show that at large times it is given in terms of the Green functions. The two time correlations can be analogously written down as, 

		\eqa{C_{\rm eq}(x,t,y,t')  =&2 \bar{\kappa} \int_{-\infty}^t ds ~\int_{-\infty}^{t'} ds' \blang \braket{x}{\nabla G_{t-s}}{BT\eta_s}  \braket{\eta_{s'} T B^\dagger }{\nabla G_{t'-s'}}{y} \brang ,}
		%
		%
		%
		%
		Taking $t > t'$, and performing the $t$ integral we have,
		\eqa{
			C_{\rm eq}(x,t,y,t')    =& ~  -2 \bar{\kappa} T^2  \int_{-\infty}^{t'} ds ~ \braket{x}{ G_{t-s}\nabla BB^{\dagger}\nabla G_{t'-s} } {y} \HTheta (t-t') , \nn }
		where $\HTheta(t)$ is the Heaviside theta function. Proceeding as before and identifying $-\bar{\kappa} \nabla BB^{\dagger}\nabla = \flap$ and interchanging $x,y$
		\eqa{C_{\rm eq}(x,t,y,t') =&  ~T^2  \int_{-\infty}^{t'} ds \left( \braket{x}{ G_{t-s}\flap G_{t'-s} }{y} +  \braket{y}{ G_{t-s}\flap G_{t'-s} }{x} \right),\nn \\
			=&~ T^2  \int_{-\infty}^{t'} ds \p_s   \braket{ x}{G_{t-s}G_{t'-s} }{y} 
			= ~T^2 G^{xy}_{t-t'}\HTheta(t-t').
		}
		Along with a similar term for $t<t'$, we can write the two time correlations as,
		\eqa{C_{\rm eq}(x,t,y,t') = \bra{x} T^2 G_{t-t'} \HTheta (t-t') + T^2 G_{t'-t} \HTheta (t'-t) \ket{y} \label{eq:twotimeenergycorr}.}

		\subsection{Current fluctuations in equilibrium}\label{sec:GreenKubo}
		Here we define the fluctuating current in the system and then establish Green-Kubo relation for the system connecting the equilibrium current fluctuations and non-equilibrium current in the system. We expect that since the total energy in the isolated system is conserved, the energy flow across the system must be in continuity form $\p_t e(x,t) + \p_x j(x,t) = 0 $. Along with the definition of current in Eq.~\eqref{eq:current}, the fluctuating current operator is defined as,    
		\eqa{\ket{ j_t} = \bar{\kappa} \fcurr \ket{e_t} - \sqrt{2\bar{\kappa}} \ket{B T \eta_t} . \label{eq:fluccurrent}}
		From previous section, it follows that the definition of current operator as $\mb{A} = -BB^\dagger \nabla$. We also note that since the current operator is odd in derivatives, the adjoint current operator has the property, $\mb A^\dagger = -\mb A$.
		Now we expect that  \cite{Kundu2009} the second moment of equilibrium total current fluctuations is related to the current in NESS through the Green-Kubo formula. A precise statement is:
\eqa{ 
\lim_{\tau \to \infty} \frac{\la q^2 \ra_{\delta T=0}}{2\tau T^2 } = \lim_{\delta T \to 0} \frac{j}{\delta T}, \label{eq:GKformulaA}
} 
		where $q(\tau) = \int_{0}^{\tau}dt \int_0^1 dx j(x,t) $. 
		In order to verify this relation, we first express  $\la q^2 \ra$, in terms of the integrals of the unequal time current correlations:
		\eqa{ \frac{\la q^2 \ra_{\delta T=0}}{\tau } =& \frac{1}{\tau } \int_{0}^{\tau} dt  \int_{0}^{\tau} dt' \int_0^1 dx \int_0^1 dy \blang j(x,t) j(y,t')\brang .\label{eq:GKformula}
		}
		\\
		Using Eq.~\eqref{eq:fluccurrent} the current correlations can be split into four parts:
		\eqa{ \big\la j(x,t) j(y,t')\big\ra =& \underbrace{\bar{\kappa}^2 \blang \bra{x} \fcurr \projop{e_t}{e_{t'}} \fcurr^\dagger \ket{y}\brang}_{\RN{1}}  + \underbrace{2 \bar{\kappa} T^2 \braket{x}{B B^\dagger}{y}}_{\RN{2}}\delta(t-t') \label{eq:currentcorrtwotime} \\ -& \underbrace{\sqrt{2} \bar{\kappa}^{3/2}  \blang \braket{x} {\fcurr}{e_t}\iprod{\eta_{t'}T B^\dagger }{y} \brang}_{\RN{3}}  -\underbrace{\sqrt{2} \bar{\kappa}^{3/2} \blang \iprod{x}{BT\eta_t }\braket{e_{t'}}{\fcurr^\dagger}{y}  \brang}_{\RN{4}} .\nn }
		Part $\RN{3}$ in the above equation can be simplified to
		\eqa{ \sqrt{2} \bar{\kappa}^{3/2}  \blang\braket{x} {\fcurr}{e_t}\iprod{\eta_{t'}T B^\dagger}{y}  \brang = &~ 2\bar{\kappa}^2  \int_{-\infty}^t ds   \blang \braket{x} {\fcurr G_{t-s}} { \nabla (BT \eta_s)} \iprod{\eta_{t'}T B^\dagger}{y} \brang ,\nn \\
			=&  - T^2 2\bar{\kappa}^2 \braket{x} {(\fcurr \nabla G_{t-t'})BB^\dagger}{y}  \theta(t-t'),\nn \\
			=&  T^2 2\bar{\kappa}^2 \bra{x} \fcurr G_{t-t'} \fcurr^\dagger \ket{y} \theta(t-t'). 
		}
		Similarly part $\RN{4}$ is given by
		\eqa{ \sqrt{2} \bar{\kappa}^{3/2}  \blang\iprod{x}{BT\eta_t }\braket{e_{t'}}{\fcurr^\dagger}{y}  \brang = &~ 2\bar{\kappa}^2 T^2 \int_{-\infty}^{t'} ds  \blang \iprod{x}{ B  \eta_{t}}  \braket { \eta_s (\nabla B)^\dagger } { G_{t'-s} \fcurr^\dagger}{y}  \brang ,\nn \\
			=&  -T^2 2\bar{\kappa}^2 \braket{x} {BB^\dagger \nabla G_{t'-t} \fcurr^\dagger}{y}  \theta(t'-t),\nn \\
			=& T^2  2\bar{\kappa}^2 \braket{x} {\fcurr  G_{t'-t}\fcurr^\dagger}{y}  \theta(t'-t),\nn 
		}
		while part $\RN{1}$, on using \eqref{eq:twotimeenergycorr}, gives
\eqa{
\bar{\kappa}^2  \blang \bra{x} \fcurr \projop{e_t}{e_{t'}} \fcurr^\dagger \ket{y}  \brang=T^2   \bar{\kappa}^2\left[ \bra{x} \fcurr   G_{t-t'}\fcurr^\dagger \theta (t-t') +  \fcurr G_{t'-t}  \fcurr^\dagger \theta (t'-t) \ket{y}\right].
}
We see that $\RN{3}+\RN{4} =  2 \RN{1}$. The first term explicitly gives,
\eqa{
  \RN{1}=& \bar{\kappa}^2   \int_0^1 dx'   \int_0^1 dy' \fcurr(x,x')\blang e_t(x')e_{t'}(y') \brang  \fcurr^\dagger(y',y), \nn \\
			=& \bar{\kappa}^2 T^2 \int_0^1 dx' \int_0^1 dy' \fcurr(x,x') \fcurr(y,y')G_{|t-t'|}(x',y'), \nn \\
			=& \bar{\kappa}^2 T^2 \sum_{n,l,l'} \hat{\mzeta}_{nl}\hat{\mzeta}_{nl'} (\lambda_{l}     \lambda_{l'})^{1/4} e^{-\bar{\kappa} \mu_n |t-t'|}    \sina_l(x)    \sina_{l'}(y).}
		Therefore,   the contribution of the parts $\RN{1}-\RN{3}-\RN{4}=-\RN{1}$ in \eqref{eq:GKformula} gives, after doing the space and time integrals:
		\eqa{ \int_{0}^{\tau}dt'  \int_{0}^{\tau}dt \int_0^1 dx  \int_0^1 dy ~(-\RN{1}) = 
			-16\bar{\kappa}^2 T^2  \sum_n \sum_{ll' \; odd}\frac{1}{ \bar{\kappa} \mu_n} \left[ \tau +\frac{(e^{-\mu_n \tau} -1)}{\mu_n}\right] \hat{\mzeta}_{nl}\hat{\mzeta}_{nl'}   (\lambda_{l}     \lambda_{l'})^{-1/4}.
		}
		On  using \eqref{eq:BBdagger}, the contribution of part  $\RN{2}$ in \eqref{eq:GKformula} gives
\eqa{
\int_{0}^{\tau}dt'  \int_{0}^{\tau}dt \int_0^1 dx  \int_0^1 dy (\RN{2}) =&  2 \bar{\kappa} T^2 \int_{0}^{\tau}dt'  \int_{0}^{\tau}dt  \int_0^1 dx  \int_0^1 dy  BB^\dagger(x,y)\delta(t-t'), \nn \\
			=& ~2 \bar{\kappa} T^2 \tau  \int_0^1  dx \int_0^1 dy \sum_{n} \frac{\alpha_n(x) \alpha_n(y)}{ \lambda_n^{1/4}}, \nn\\
			=& 16 \bar{\kappa} T^2  \tau \sum_{n \; odd} \frac{1}{ \lambda_n^{5/4}}.
}
Combining the above results we finally have 
\eqa{ 
\lim_{\tau \to \infty} \frac{\la q^2 \ra_{\delta T=0}}{2\tau T^2 } = \bar{\kappa} \left( 8\sum_{n \; odd} \frac{1}{ \lambda_n^{5/4}} - \sum_n \sum_{ll' \; odd}\frac{8}{ \mu_n} \frac{\hat{\mzeta}_{nl}\hat{\mzeta}_{nl'}  }{  (\lambda_{l}     \lambda_{l'})^{1/4}}\right) .  \label{eq:currfluc}
}

The first summation yields $0.5050\ldots$  and the second yields $\approx 0.0931$, hence we get
\eqa{
\lim_{\tau \to \infty}  \frac{\la q^2 \ra_{\delta T=0}}{2T^2\tau} \approx 0.4119\bar{\kappa} , 
} 
which, up to numerical accuracy is consistent with the numerical value of steady state current ($j/\delta T = 0.4124\bar{\kappa}$) we found in \eqref{eq:sscurrent}, thus validating the Green-Kubo formula in \eqref{eq:GKformulaA}. 

Note that in order to get the expected scaling in system size $N$, we need to put in the appropriate length scaling of the eigenvalues and eigenfunctions, for example  $\lambda_n \to \lambda_n/N^2$ and $\mu_n \to \mu_n /N^{3/2}$). We also need to consider the integrated current $Q(\tau) = \int_{0}^{\tau}dt \int_0^N dx j(x,t) $ and then one gets
 $\lim_{\tau \to \infty}  \frac{\la Q^2 \ra_{\delta T=0}}{2T^2\tau} \approx \frac{ 0.4119\bar{\kappa}}{\sqrt{N}}$ and $\frac{J}{\delta T} = \frac{j}{ \sqrt{N} \delta T}\approx \frac{0.4124\bar{\kappa}}{\sqrt{N}}$.

		\begin{figure}
			
			\centering
			\includegraphics[width=0.75\linewidth]{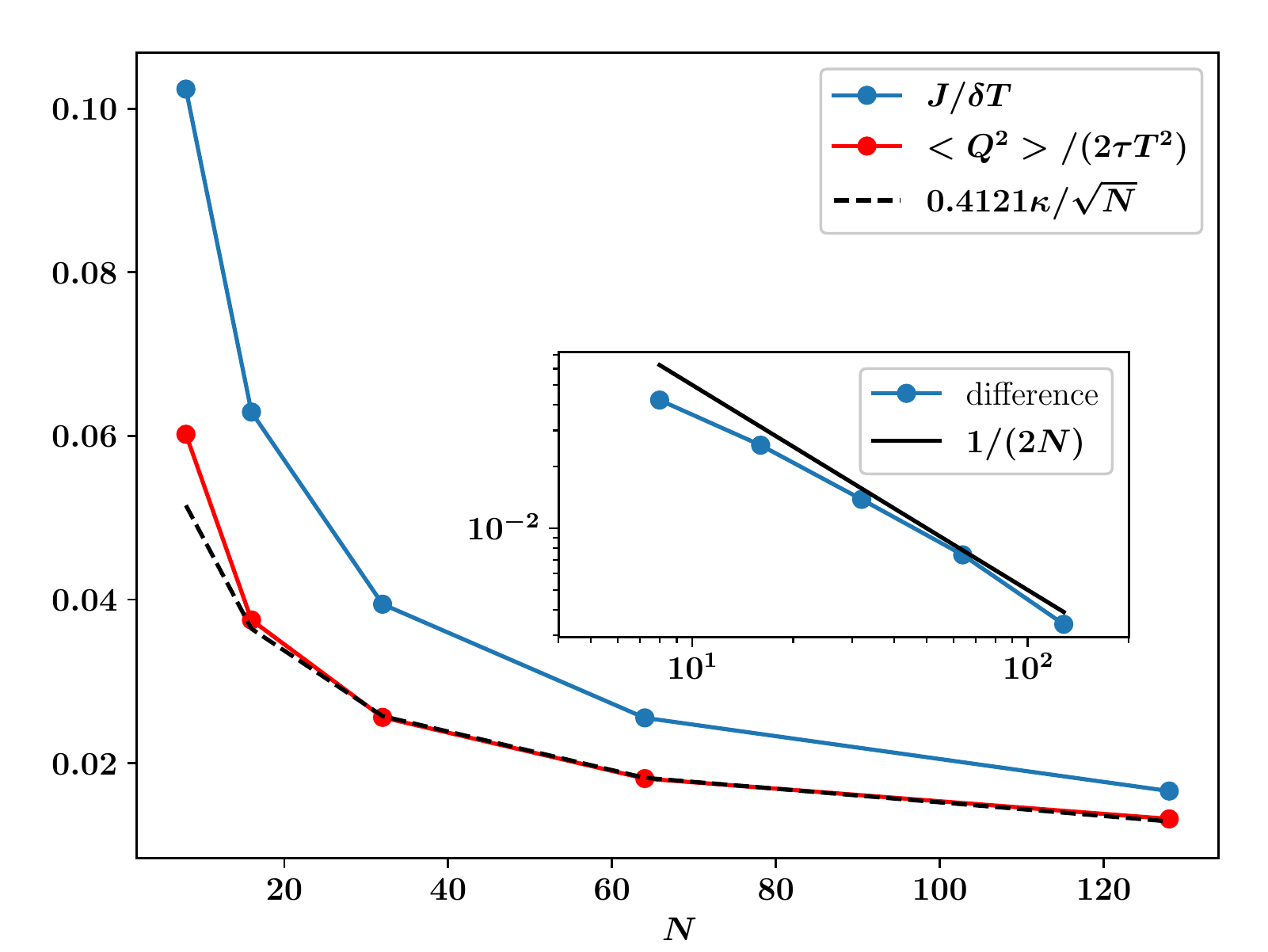}
			\caption{For the microscopic HCME model, we compute the two quantities,  $J/\delta T$ computed from non-equilibrium simulations connected to heat baths and $ <Q^2>/(2 \tau T^2)$ computed from equilibrium simulations, are plotted as a function of $N$. The black dashed curve is for the theoretical current with appropriate scaling as given in Eq.~\eqref{eq:sscurrent}. We find that for small $N$, these two do not match, and the difference between the two decays as $1/N$ (inset), which is due to the contribution of current from the stochastic part. This signifies that at large $N$, Green-Kubo holds while for small $N$, it fails.}
			\label{fig:greenkubo}
		\end{figure}

{ The above verification of the Green-Kubo identity was obtained using the fluctuating fractional diffusion equation, which is valid in the limit of large system size. A natural question is as to whether the identity is true even for a small chain with the microscopic dynamics (HCME), as would be expected from the fluctuation theorem.  In Fig.~\eqref{fig:greenkubo}, 
we present a numerical comparison of the equilibrium current fluctuations, with  the non-equilibrium current,  both computed from the microscopic model for finite systems. We see clear evidence that for small $N$, the  Green-Kubo relation is violated in the HCME model. We also find that the difference between the fluctuation and response parts decreases with system size as $\sim 1/N$. Somewhat surprisingly, the numerically obtained fluctuations (from HCME simulations) are very close to the response computed from the fractional diffusion  equation description. A possible reason for the failure of the fluctuation theorem for small systems could be that in this model, 
the Hamiltonian part of the current (which goes as $1/\sqrt{N}$), and the stochastic part of the current ($\sim 1/N$) have different time-reversal symmetries.}  
		\subsection{General fractional power}
		In this section, we discuss a possible generalization of  the  results of the previous section for the Green-Kubo identity to the case of arbitrary fractional power $\genpower$ of the Laplacian. There is currently no known microscopic model in which heat transfer can be described by a fractional equation with arbitrary $\beta$ --- nevertheless it is an interesting exercise as it leads to some general mathematical identities involving Riemann-zeta functions. Using the definition of fractional Laplacian in Eq.~\ref{Lbeta}, namely through the operation $\flapfinite^{(\genpower)}  \phi_n(x) =  \lambda_n^{\genpower} \phi_n(x)$, we can proceed in a similar way as for the $\beta=3/4$ case and compute steady state properties in NESS as well as equilibrium current fluctuations.

Corresponding  to  Eq.~\eqref{eq:sscurrent}  we then get
\eqa{
\frac{j}{\delta T}= \frac{1}{8 \left(2^{2\beta}-1 \right) (2\pi) ^{-2\beta } \zeta (2\beta )}, \label{eq:gennessj}
}
and corresponding to Eq.~\eqref{eq:currfluc} we get
\eqa{
\lim_{\tau \to \infty} \frac{\la q^2 \ra_{\delta T=0}}{2\tau T^2 } = 8  \left(1-2^{2 \beta -4}\right) \pi ^{2 \beta -4} \zeta (4-2\beta ) - \sum_{n \; even} \sum_{ll' \; odd}\frac{8}{ \mu_n^{(\genpower)}} \frac{ \hat{\mzeta}_{nl}^{(\genpower)}\hat{\mzeta}_{nl'}^{(\genpower)} }{  (\lambda_{l}     \lambda_{l'})^{1-\genpower}}, 
\label{eq:genjfluc}}
where due to structure of $\hat{\mzeta}^{(\beta)}_{n,l}$, only the terms with even $n$ survives for odd $l$. This is computed as before but now with power $\genpower$ and is explicitly given as
\eqa{
  \hat{\mzeta}^{(\beta)}_{2k+2,2m+1} &=\frac{D^{(\beta)}_{2k+2}}{\lambda_{2m+1}^{\beta} -\mu^{(\beta)}_{2k+2}} , ~~~ k,m \ge 0~, \\
{\rm where}~~  D^{(\beta)}_{2k+2} &= \left[\sum_{m\ge0} \frac{1}{(\lambda_{2m+1}^{\beta}-\mu^{(\beta)}_{2k+2})^2}\right]^{-1/2},
}
and $\{\mu_{2n}^{(\beta)}\}_{n \ge 1}$ are the ordered roots of the equation
\eqa{
  \sum_{k\ge0}  \frac{1}{\lambda_{2k+1}^{\beta} - \mu^{(\beta)}} = 0~.
} 
All the coefficients in the above expressions are explicit and we have evaluated numerically the right hand sides of Eqs.~(\ref{eq:gennessj},~\ref{eq:genjfluc}) 
for values of $\genpower \in (0.5,1.5)$. In Fig.~\eqref{fig:genfrac}, we plot  these quantities  and find that they are very close to each other, hence verifying the Green-Kubo formula Eq.~(\ref{eq:GKformulaA}) for general $\beta$.
The differences  arise from  numerical error due to truncation of series and also use of a finite number of basis functions. For $\beta=1$, this leads to diffusive results for which the double summation can be computed explicitly. 
Conversely, on the basis of the validity of the Green-Kubo formula we are then led to conjecture a mathematical identity between the right hand sides of Eqs.~(\ref{eq:gennessj},~\ref{eq:genjfluc}). 
For $\beta < 1/2$, one has a non-convergent series summation in Eq.~\eqref{eq:genjfluc}, which leads to a breakdown of the identity in this form. This corresponds to defining zeta function for power less than $1$,  and possibly analytic continuation could extend the definition to other values of $\beta$. 
We believe that the relation holds true at least in the open interval $\genpower \in (1/2,3/2)$. However, proving it remains an open problem.

\begin{figure}
\centering
\includegraphics[width=0.75\linewidth]{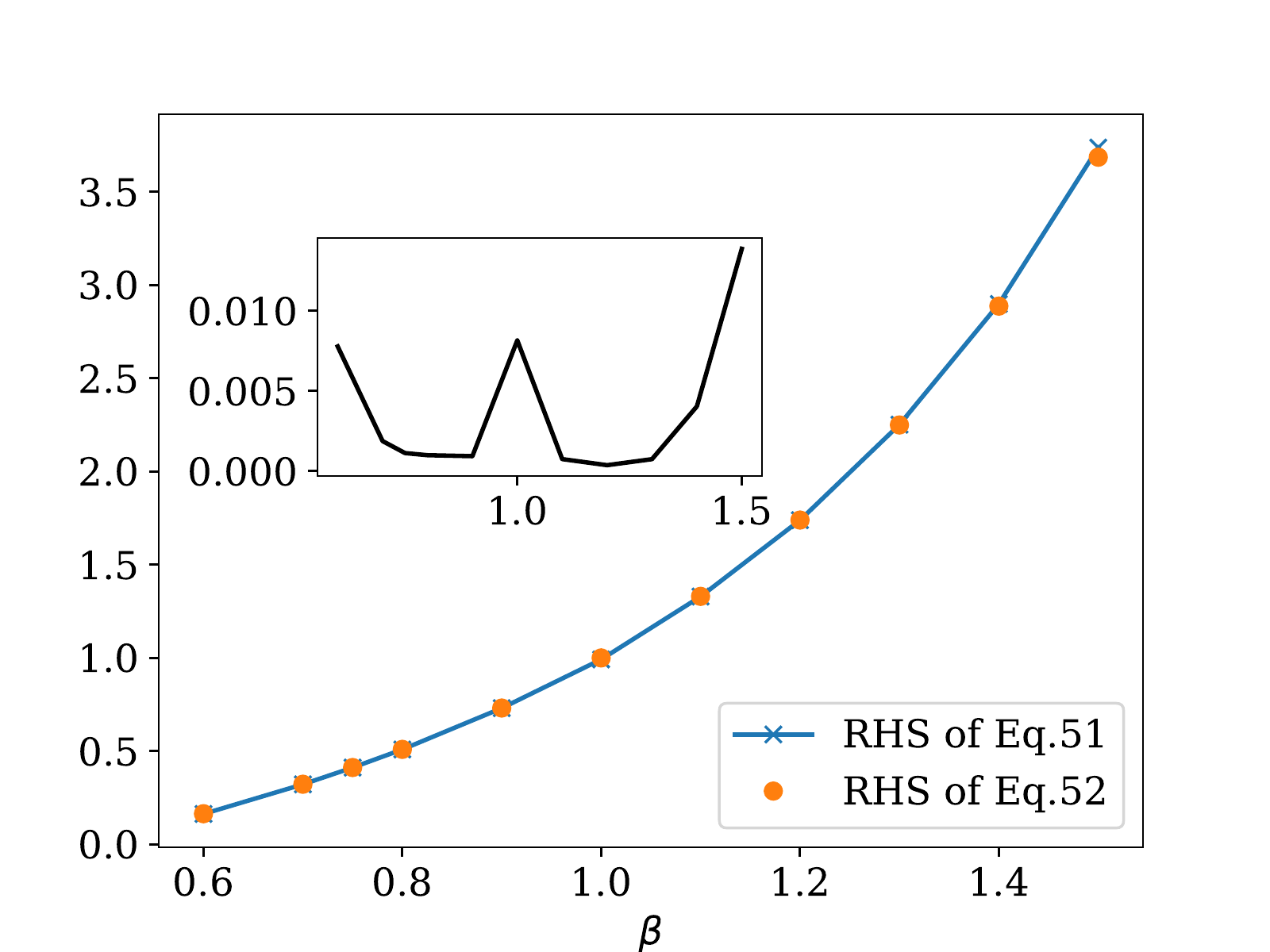}
\caption{The  numerically evaluated expressions in Eq.~\eqref{eq:gennessj} and Eq.~(\ref{eq:genjfluc}) are plotted as a function of general fractional power $\beta$. The two quantities match numerically to a very good precision. The relative error between the two is plotted in the inset.}
\label{fig:genfrac}
\end{figure}

\subsection{Long range correlations in NESS}\label{sec:Longrange}
 For a noneqlibrium current carrying steady state, it is expected that fluctuations across the system will develop non-zero long-range correlations. These long-range correlations is a distinguishing feature of non-equilibrium systems with conservative dynamics \cite{Garrido1990}.
		 In some diffusive  lattice gas  as well as some Hamiltonian systems, these long-range correlations have been studied \cite{Spohn1983,Bodineau2008,Bertini2007,Derrida2007,Bernardin2012a}. 
		The energy correlation in the velocity flip model (HCMF) in NESS is defined as $C_{NESS}(x,y) = \la e(x,t) e(y,t) \ra $, where the average is taken in NESS (as $t \to \infty $). It was shown that  $C_{NESS}(x,y) = \delta T ^2 \Delta ^{-1} (x,y)$, where $\Delta$ is the Laplacian operator with Dirichlet boundary conditions. 
		From the definition of fluctuating fractional equation in equilibrium, it is tempting to extend the definition of fluctuating fractional equation to non-equilibrium case, where the temperature is space-dependent: 
		\eqa{
			\p_t \ket{e_t} &= -\flap \ket{e_t} + \sqrt{ 2 \bar{\kappa}} \nabla (B T_{NESS} \ket{\eta_t}),
			\label{eq:fluctuatingeqneq}
		}
		We note that there is an ambiguity regarding the relative position of the operator $B$ and $T_{NESS}$, and also with the   definition for operator $B$ and $B^\dagger$ separately. If we anyway proceed with a naive replacement of $T $ by $T_{NESS}(x)$ in Eq.~\eqref{eq:fluctuatingeq}, to get \eqref{eq:fluctuatingeqneq}, we can perform the computation of $C_{NESS}(x,y)$ and find that this does not agree with the results from direct simulations.  However, in analogy to the HCMF model, we conjecture that the NESS energy correlations $C_{NESS}(x,y)$ are given (upto a constant factor $\nu$) by the inverse of the fractional Laplacian (in Dirichlet basis):  
		\eqa{\delta T^2 \mb{C}(x,y) = \frac{ \delta T^2}{\nu} \mb{L}^{-1} = \frac{ \delta T^2}{\nu}\sum_{n\ge 1} \frac{\psi_n(x)\psi_n(y)}{\mu_n}, \label{eq:lrcorr} }
where we have defined $\delta T^2 \mb{C}(x,y)=C_{NESS}(x,y)-T_{NESS}(x)^2\delta(x-y)$, with the local same-site correlation $T_{NESS}(x)^2\delta(x-y)$ subtracted from correlations.

		{\bf Numerical verification of Eq.~\eqref{fig:LongRangeCorrelations}}:		We simulate the microscopic system in non-equilibrium with two Langevian heat baths kept at different temperature. After the system is in the steady state, we compute $\mb{C}(x,y) =   N \la e(i/N)e(j/N)\ra $ where $e(i/N)=E(i/N)-\la E(i/N)\ra$. In Fig.~\eqref{fig:LongRangeCorrelations} we compare our conjectured form from Eq.~ \eqref{eq:lrcorr} with the results from microscopic simulations. We see that with the constant  $\nu \approx 3.77$, the two numerical curves (for $y=1/4$ and $y=1/2$) match well with the inverse of fractional Laplacian. The constant $\nu$ is related to the total energy fluctuations in the system at NESS as
		\eqa{\int_0^1 dx \int_0^1 dy \frac{ \delta T^2}{\nu} \mb{L}^{-1}(x,y)  + \int_0^1 dx T^2_{NESS}(x) =  \int_0^1 dx \int_0^1 dy ~ C_{NESS}(x,y)=\blang \Delta E_{tot}^2\brang_{NESS}.  \nonumber 
		}
By evaluating the integrals on the LHS and finding the RHS from numerical simulations in the NESS, we can use the above equation to independently evaluate $\nu$. We find that the  fluctuations $\blang \Delta E_{tot}^2\brang_{NESS}$,  obtained from   simulations in NESS, converges very slowly  and with the final accessed simulation time {($2 \times 10^9$ time with $10^8$ samples)} we estimate $3.51 \le \nu \le 4.2$. The value $\nu \approx 3.77$, obtained by fitting the long range correlations data from simulation, is well within the limits of the above estimate.
		 We have tested (see \ref{app:longrangetests}) that the constant $\nu$ does not change substantially with $\delta T$ and $\bar{T}$, within the numerical accuracy and finite size effects.  We note in \ref{app:longrangetests}, that if we did the same computation with $\sin(n \pi x)$ basis, then the results would differ significantly. 		 
		 We close this section by making a comment that proving our conjecture on the equality between the long-range correlations and the inverse fractional operator is an open question.
		
		\begin{figure}
			\centering
			\includegraphics[width=0.45\linewidth]{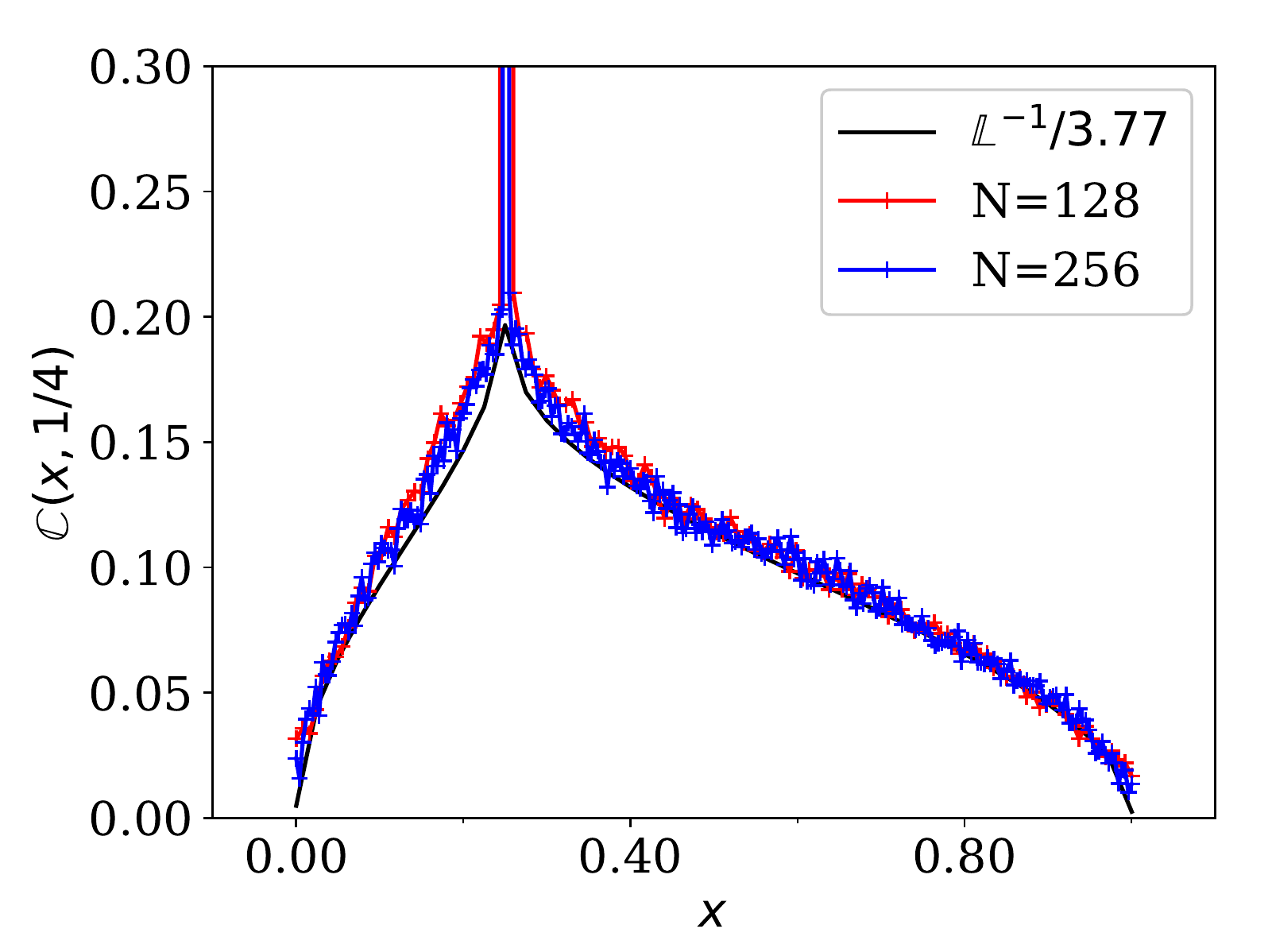}
			\includegraphics[width=0.45\linewidth]{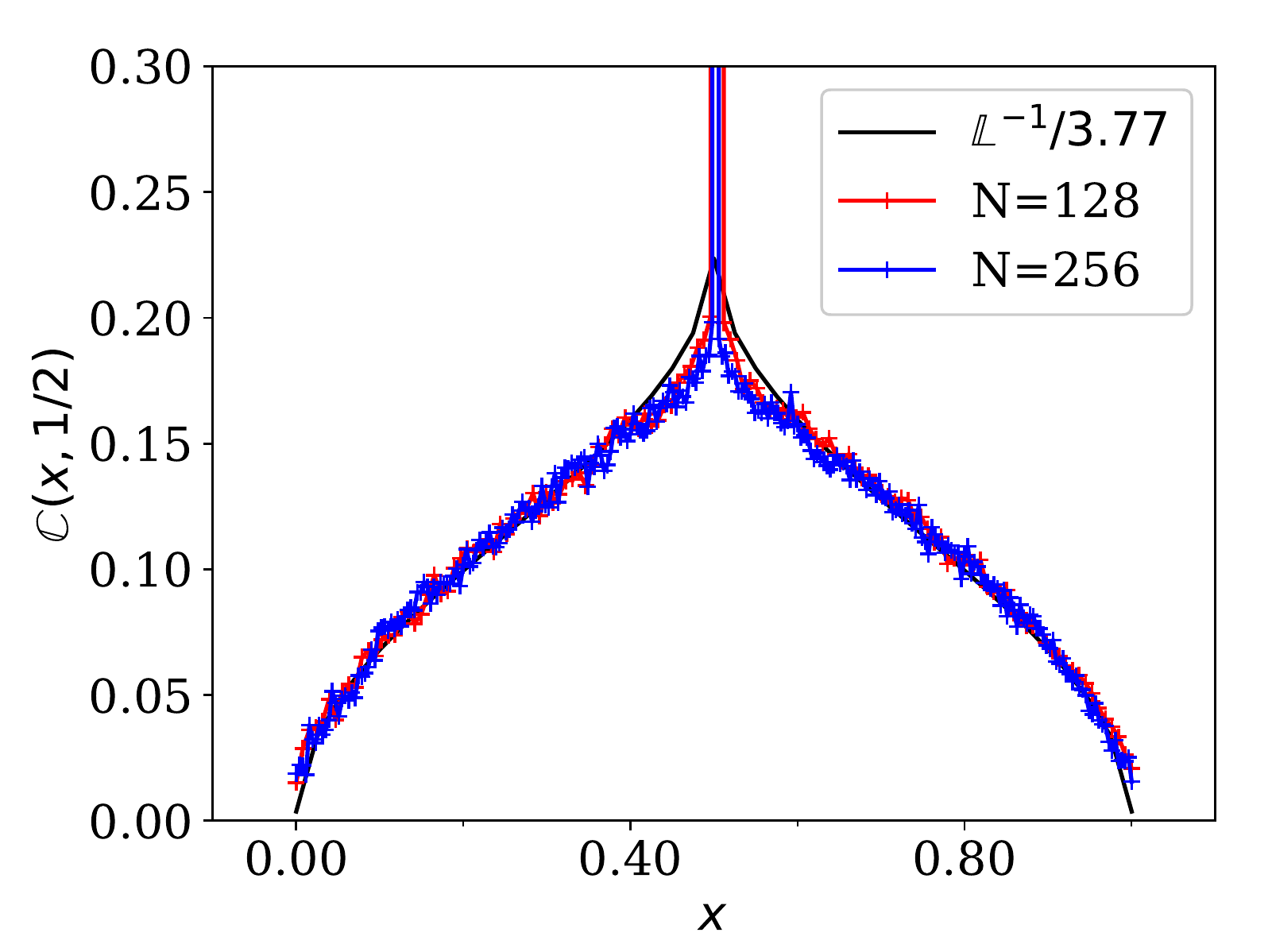}
			\caption{Non-equilibrium energy correlation function $\mb{C}(x,y) $ in steady state of Harmonic chain momentum exchange model for $y=1/4$ (left figure) and $y=1/2$ (right figure). The system size considered here are for $N=128,256$ with $T_L=2, T_R=1$ . The inverse of the fractional Laplacian (summed up to 600 basis states) and  with an arbitrary constant factor ($\nu=3.77$), is plotted (black solid) along with the simulation results. }
			\label{fig:LongRangeCorrelations}
		\end{figure}

		
		\section{Conclusions}
		\label{sec:conc}
		

		We have shown that in a particular analytically tractable model of heat conduction in one dimension, the macroscopic evolution of energy in an open system is governed by the fractional diffusion equation. This gives us a definition of the fractional operator in a finite domain and also gives a meaning to the fractional operator in terms of linear PDE's (similar to the harmonic extension of a fractional operator). We describe an efficient procedure to numerically construct the eigenspectrum of this operator. In terms of this operator, we compute the steady state and time evolution of temperature field, which we compare with microscopic simulations of the system. We defined the fluctuating fractional equation and used it to verify the Green-Kubo relation in the system. We also generalize the Green-Kubo for general fractional power which leads to some general mathematical identity involving zeta functions. This identity is verified numerically. We also conjecture that the long-range correlations are given by the inverse of a fractional operator. Proving this conjecture in Eq.~\eqref{eq:lrcorr}, as well as finding the correct equation to replace Eq.~\eqref{eq:fluctuatingeqneq}  are open problems. The other interesting question would be to consider the use of the fractional operator in studying the dynamics of other Hamiltonian systems 
such as the FPUT model and also the HCME model with other boundary conditions. Another very interesting aspect is to study the usefulness of the eigensystem of the fractional operators in studying other applications where the underlying dynamics can be modelled as Levy flights or Levy walks.


		
		\section{Acknowledgments}
		
		
		The authors are very grateful to G. Basile, T. Komorowski and S. Olla to send them their unpublished notes \cite{PrivateCommunication2015} on which a part of this work is based.  Aritra Kundu would like to thank the hospitality of Nice Sophia-Antipolis University Laboratoire Dieudonn\'e and Institut Henri Poincar\'e - Centre Emile Borel during the trimester ``Stochastic Dynamics Out of Equilibrium" where part of the work was done. This work benefited from the support of the project EDNHS ANR-14-CE25-0011 of the French National Research Agency (ANR), and also in part by the International Centre for Theoretical Sciences (ICTS) during a visit for participating in the program Non-equilibrium statistical physics (Code: ICTS/Prog-NESP/2015/10). C\'edric Bernardin thanks the French National Research Agency (ANR) for its support through the grant ANR-15-CE40-0020-01 (LSD) and the European Research Council (ERC) under  the European Union's Horizon 2020 research and innovative programme (grant agreement No 715734). Anupam Kundu (AK) would like to acknowledge the support from DST grant under project No. ECR/2017/000634. AK, AD and CB would like to acknowledge the support from the project 5604-2 of the Indo-French Centre for the Promotion of Advanced Research (IFCPAR). KS was supported by JSPS Grants-in-Aid for Scientic Research (JP16H02211 and
		JP17K05587).

		
		\section{References}
		%
		%

		\providecommand{\newblock}{}

		\appendix
				\section{Connection coefficients between sine and cosine}\label{app:cossin}
		We can expand $\sin$ in the complete basis of $\cos$ as,
		$\alpha_n(y) = \sum_l \mc{T}_{nl} \phi_l(y)$, with explicit coefficients $\mc{T}_{nl}=\int_0^1 dy \alpha_n(y) \phi_l(y)$. The coefficients are given as,
		\eqa{
			\mc{T}_{nl}   =&  \begin{cases}
				0 ~\text{if l=0, n is even,}\\
				\frac{2\sqrt{2}}{\pi n} ~\text{if l=0, n is odd,}\\
				\frac{2}{\pi}\left[\frac{\delta_{(n+l),\text{odd}}}{n+l}+\frac{\delta_{(n-l),\text{odd}}}{n-l}\right],~\text{if}~ l > 0.
			\end{cases}
		}
		\section{Derivation of matrix equations of Fractional operator}\label{app:matrixrep}
	Here we enumerate the steps involved in going from the set of PDE's to the matrix representation of $\mb L$ as stated in the main text. The correlation and temperature fields are expanded as, 
		$
			C(x,y,\tau) -C_{ss}(x,y)= \sum_{n=1}^\infty \hat{C}_n(x,\tau) \alpha_n(y)$ and $
			T(y,\tau) -T_{ss}(y)= f(y,\tau)=\sum_{n=1}^\infty \hat{f}_n(\tau) \alpha_n(y) .
		$
		Following \cite{Lepri2010}, the first of the equations in \eqref{PDE} implies  
		$\p^4_x\hat{C}_n(x) = -4 \delta_n^4 \hat{C}_n(x)$, where  $\delta_n = \sqrt{n \pi \omega/(2\gamma)}$. Solving these equations with the appropriate boundary conditions one eventually gets,
		\eqa{
			\hat{C}_n(x,\tau)= \hat{A}_n(\tau) e^{-\delta_n x} ~[\sin (\delta_n x)-\cos (\delta_n x)] ~.
		}
	using the PDE's one gets
	\eqa{ 	
		\hat{A}_n(\tau) &= -\frac{1}{4\gamma \delta_n} \sum_{k=1} \mc{T}_{kn}^\dagger \sqrt{\lambda_k} f_k \nn	\\
		\dot{\hat{f}}_m  &= \omega^2 \sum_{n=1} \mc{T}_{mn}  \sqrt{\lambda_n} \hat{A}_n(\tau) \nn\\
		=&  -\frac{\omega^2}{4\gamma} \sum_{n,k=1} \mc{T}_{mn}  \sqrt{\lambda_n} \frac{1}{\delta_n}   \mc{T}_{kn}^\dagger \sqrt{\lambda_k} f_k \nn\\
		=&  -\frac{\omega^2}{4\gamma} \sum_{n,k=1} \mc{T}_{mn}   \frac{\lambda_n}{\delta_n}   \mc{T}_{nk}^\dagger f_k \nn\\
		=& -\bar{\kappa} \left[\mc{T} \Lambda^{3/4}\mc{T}^\dagger\right]_{mk} f_k
}
where we used the property of transformation element, $\mc{T}_{kn}^\dagger \sqrt{\lambda_k} = \mc{T}_{kn} \sqrt{\lambda_n}$.
		with $\mc{T}_{nl}=\iprod{\alpha_n}{\phi_l}=\int_0^1 dy \alpha_n(y) \phi_l(y)$ and the constant $\bar{\kappa}$.
	
		\section{Formal identities of cos and sin series}\label{app:cossumidentity}
		Consider the two Fourier cosine series on $[0,1]$, 
		\eqa{
			q =&~   \frac{1}{2} + \sum_{ m \; odd} \frac{-2 \sqrt{2} }{\pi^2m^2}\sqrt{2}\cos(\pi m q),\nn\\
			q^2 -q =& -\frac{1}{6} + \sum_{m \in even } \frac{2 \sqrt{2}}{\pi^2m^2}\sqrt{2} \cos(\pi m q),\\
			q^2 - q =& \sum_{m \in odd} -\frac{4\sqrt 2}{(n \pi)^3} \sqrt{2} \sin(\pi m q).\nn
		}
		Formally differentiating these two equations with respect to $q$ on both sides we get two formal identities (which needs to be interpreted as distributional sense): 
		
		\eqa{
			\sum_{ m \;  odd} cos(\pi m q) =& ~0\\
			\sum_{\substack{ m \;  even, \\m>1}} \sqrt{2} cos(\pi m q) =&  -\frac{1}{\sqrt{2}},\\
					\sum_{ m \;  odd} \frac{\sqrt{2}\sin(m \pi  q)}{\sqrt{\lambda_n}}=&                                                                                                                                              \frac{1}{2\sqrt{2}}.
			\label{identity1}
		}

		\section{Alternate series representation of $BB^\dagger$ and connection with Eq.~\eqref{eq:BBdagger} \label{app:BBdaggergen}}
		As mentioned in the main text the kernel operator $BB^\dagger$ has appeared earlier in the context of heat conduction through HCME model \cite{Cividini2017}. Using non-linear hydrodynamics theory in \cite{Cividini2017}, the non-local linear response relation has been established for general boundary conditions characterized by a reflection coefficient $R=\left( \frac{\lambda-\omega}{\lambda+\omega}\right )^2$ which vary from $0$ to $1$. 
		For given $R$, the expression for the kernel is given as \cite{Cividini2017}
		\begin{align}
		BB^\dagger(x,x')=\frac{1}{\sqrt{2\pi}} 
		\sum_{n=-\infty}^{\infty} \left[ \frac{R^{|2n|}}{\sqrt{|2n+x-x'|}}-\frac{R^{|2n+1|}}{\sqrt{|2n+x+x'|}}\right].
		\end{align} 
The value $R=0$ corresponds to the resonance condition $\omega=\lambda$ for free boundary condition \emph{i.e.} $q_0=q_1$ and $q_{N}=q_{N+1}$.  On the other hand,  $R=1$ corresponds to fixed boundary condition. For $R=1$, one can explicitly check with the above representation that 
		\eqa{
		\int_0^1dx'~BB^\dagger(x,x')\alpha_m(x')=\frac{1}{\lambda_{n}^{1/4}}\alpha_m(x),
		\label{eq:sintransformation}}
which is same as Eq.~\eref{eq:BBdagger}. The proof is as follows. The LHS of Eq.~\eqref{eq:sintransformation} can be written as,
\begin{align}
	L.H.S=\frac{1}{\sqrt{\pi} }\int_0^1  dy  \Bigg[ \frac{1}{\sqrt{|x-y|}} &+\sum_{n=1}^\infty \frac{1}{\sqrt{2n+x-y}} + \frac{1}{\sqrt{2n-x+y}} \nonumber \\ 
	&-\frac{1}{\sqrt{2n-2+x+y}} - \frac{1}{\sqrt{2n-x-y}}  \Bigg ] \sin(m \pi y)
\end{align}
	Using change of variables  and separating the part in absolute value we have,
\eqa{L.H.S=\frac{1}{\sqrt{\pi} } \Bigg[ \int_0^x  dz &\frac{\sin(m\pi (x-z))}{\sqrt{z}}+\int_0^{1-x}  dz \frac{\sin(m\pi (x+z))}{\sqrt{z}} \nn \\
+& \sum_{n=1}^\infty \Bigg(  \int_{2n-1+x}^{2n+x} dz \frac{\sin (m\pi (2n+x-z))}{\sqrt{z}} + \int_{2n-x}^{2n+1-x} dz \frac{\sin (m\pi (z-2n+x))}{\sqrt{z}} \nn\\
  -& \int^{2n-1+x}_{2n-2+x} dz \frac{\sin (m\pi (z-2n+2-x))}{\sqrt{z}}- \int^{2n-x}_{2n-1-x} dz \frac{\sin (m\pi (2n-x-z))}{\sqrt{z}}
		\Bigg)\Bigg] \nn}
Upon using trigonometric identities this can be reduced to,
\eqa{L.H.S.=\frac{1}{\sqrt{\pi} } \Bigg[ \Bigg( \int_0^x  dz +   \sum_{n=1}^\infty \Big(  \int^{2n-1+x}_{2n-2+x} dz +  \int_{2n-1+x}^{2n+x} \Big)\Bigg) \frac{\sin(m\pi (x-z))}{\sqrt{z}} \nn\\
 + \Bigg( \int_0^{1-x}  dz +   \sum_{n=1}^\infty \Big(  \int^{2n-x}_{2n-1-x} dz +  \int^{2n+1-x}_{2n-x} \Big)\Bigg) \frac{\sin(m\pi (x+z))}{\sqrt{z}} \Bigg]}
which, upon simplifying further provides the R.H.S. of Eq.~\eqref{eq:sintransformation},
	\eqa{\frac{1}{\sqrt{\pi} } \int_0^\infty dz  \Bigg[ \frac{\sin(m\pi (x-z))}{\sqrt{z}}+\frac{\sin(m\pi (x+z))}{\sqrt{z}} \Bigg] = \frac{1}{\sqrt{m\pi}} \sqrt{2} \sin(m \pi x)}

		\section{Explicit expressions of some equations mentioned in maintext}
		The total fractional equation \eqref{eq:fluctuatingeq} can be written explicitly as,
		\eqa{\p_t e(x,t) &= \nabla_x \int dx' \left[ {\mathbb A}(x',t) e(x',t) -  B(x',t) T \eta(x',t)\right]
		}
		The equation for the two time equilibrium spatio-temporal correlation (Eq.~  \eqref{eq:twotimeenergycorr})can be written as,
		\eqa{C_{\rm eq}(x,t,y,t')  =&2 \bar{\kappa} \int_{-\infty}^t ds ~\int_{-\infty}^{t'} ds'  \int dx'  \int dy' \nabla_{x'} G_{t-s}^{xx'} \blang  {(BT\eta)}(x',s)   (BT\eta )(y',s') \brang {\nabla_{y'} G_{t'-s'}^{yy'}}}
		The spatio-temporal current correlations in Eq.~ \eqref{eq:currentcorrtwotime} can be explicitly written as,
		\eqa{ \blang  j(x,t) j(y,t')\brang = \int dx' \int dy' \bar{\kappa}^2 \fcurr(x,x')\fcurr(y,y') \blang  e(x',t)e(y',t') \brang \label{eq:appj1} \\
			+ 2 \bar{\kappa} T^2 B(x,x')B(y,y') \blang \eta(x',t)\eta(y',t') \brang \label{eq:appj2}\\
			-\sqrt{2}\bar{\kappa}^{3/2}T \fcurr(x,x')B(y,y') \blang e(x',t)\eta(y',t')\brang \label{eq:appj3} \\
			-\sqrt{2}\bar{\kappa}^{3/2} T B(x,x')\fcurr(y,y') \blang \eta(x',t)e(y',t')\brang\label{eq:appj4} 
		}
		Since it might be a bit confusing using the symbolic vector notation for the operations in the main text, here we show explicitly the expressions for individual terms and show the 2nd and 3rd terms give a similar term to 1st.
		Eq.~ \eqref{eq:appj1} gives,
		\eqa{\bar{\kappa}^2\int dx' \int dy'  \fcurr(x,x')\fcurr(y,y') G^{x'y'}_{|t-t'|} &=\bar{\kappa}^2\int dx' \int dy'  BB^\dagger (x,x')\p_{x'} BB^\dagger (y,y')\p_{y'} G^{x'y'}_{|t-t'|}\nn\\
			=& \bar{\kappa}^2\int dx' \int dy'  \mb{A}(x,x') G^{x'y'}_{|t-t'|} \mb{A}^\dagger(y',y)
		}
		where we used the adjoint representation for $\mb{A}^\dagger(y,y') =  \mb{A}(y',y) $.
		Eq.~\eqref{eq:appj3} gives,
		\eqa{
			\RN{3} =& \int dx' \int dy'\sqrt{2}\bar{\kappa}^{3/2}T \fcurr(x,x')B(y,y') \blang e(x',t)\eta(y',t')\brang \nn  \\
		=& -2\bar{\kappa} T^2 \int dx' \int dy'  \int dx''  \int dy'' \int_{-\infty}^t ds  BB^\dagger (x,x')\p_{x'} B(y,y')G^{x'x''}_{t-s} \p_{x''} B(x'',y'')\blang \eta(y'',s)\eta(y',t')\brang \nn\\
			=&2\bar{\kappa} T^2 \int dx' \int dy' \int dx'' BB^\dagger (x,x')\p_{x'} B(y,y')\p_{x''} (G^{x'x''}_{t-t'})  B(x'',y') \theta(t-t')\nn \\
			=&2\bar{\kappa} T^2 \int dx' \int dy' \int dx'' \fcurr(x,x')  G^{x'x''}_{t-t'}  \fcurr^\dagger (x'',y)\theta(t-t').
		}
		Eq.~\eqref{eq:appj4} gives,
		\eqa{
			\RN{4}=& \int dx' \int dy'\sqrt{2}\bar{\kappa}^{3/2} T B(x,x')\fcurr(y,y') \blang \eta(x',t)e(y',t')\brang\nn\\
		 =& -2\bar{\kappa} T^2 \int dx' \int dy'  \int dx''  \int dy'' \int_{-\infty}^t ds B(x,x')\fcurr(y,y') G^{y'x''}_{t'-s} \p_{x''} B(x'',y'')\blang \eta(x',t)\eta(y'',s)\brang \nn\\
			=& 2\bar{\kappa} T^2 \int dx' \int dy' \int dx'' B(x,x')\fcurr(y,y')  \p_{x''}(G^{y'x''}_{t'-t}) B(x'',x') \theta(t'-t)\nn \\
			=&2\bar{\kappa} T^2 \int dx' \int dy' \int dx'' \fcurr(y,y')  G^{y'x''}_{t'-t}  \fcurr^\dagger (x'',x)\theta(t'-t).
		}
	The second term is explicit in the main-text.
		
		\section{Some tests on Long-range correlations}\label{app:longrangetests}
	In Fig.~\eqref{appfig:longrangedifftemp} we do microscopic simulations for different temperature differences, and absolute temperatures to show, to good accuracy, the constant $\nu$ does not depend on these factors.  In Fig.~ \eqref{appfig:longrangesin} we test the use of $\alpha$(sin) basis instead of the $\psi$ basis for theoretical prediction for the nature of long-range correlations, and show it performs badly. 
				\begin{figure}
			\centering
			\includegraphics[width=0.75\linewidth]{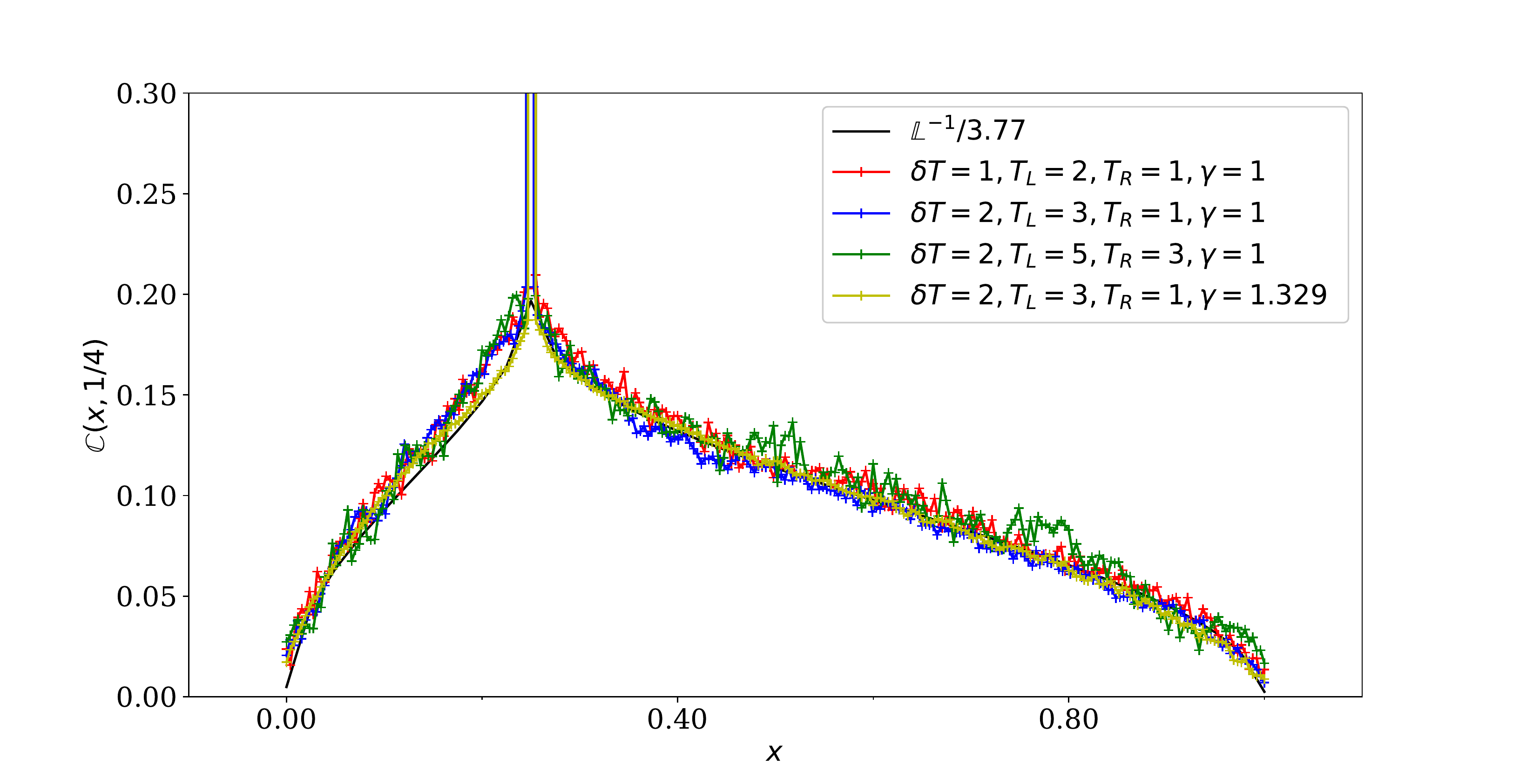}
			
			\caption{Simulation results for $N=256$, with $4$ different temperature and exchange rate parameters mentioned in the graph. The black curve is theoretically computed curve with $\nu = 3.77$, we see the results from the simulation are very close to predicted theoretical curve for the different parameters. This suggests that the parameter $\nu$ is independent of absolute value of the applied boundary temperature, the temperature gradient of the system and the long range correlations do not depend on the details like the stochastic exchange rate etc. The slight differences are again, hopefully a result of finite size effect.}    
			\label{appfig:longrangedifftemp}
		\end{figure}
		\begin{figure}
			\centering
			\includegraphics[width=0.65\linewidth]{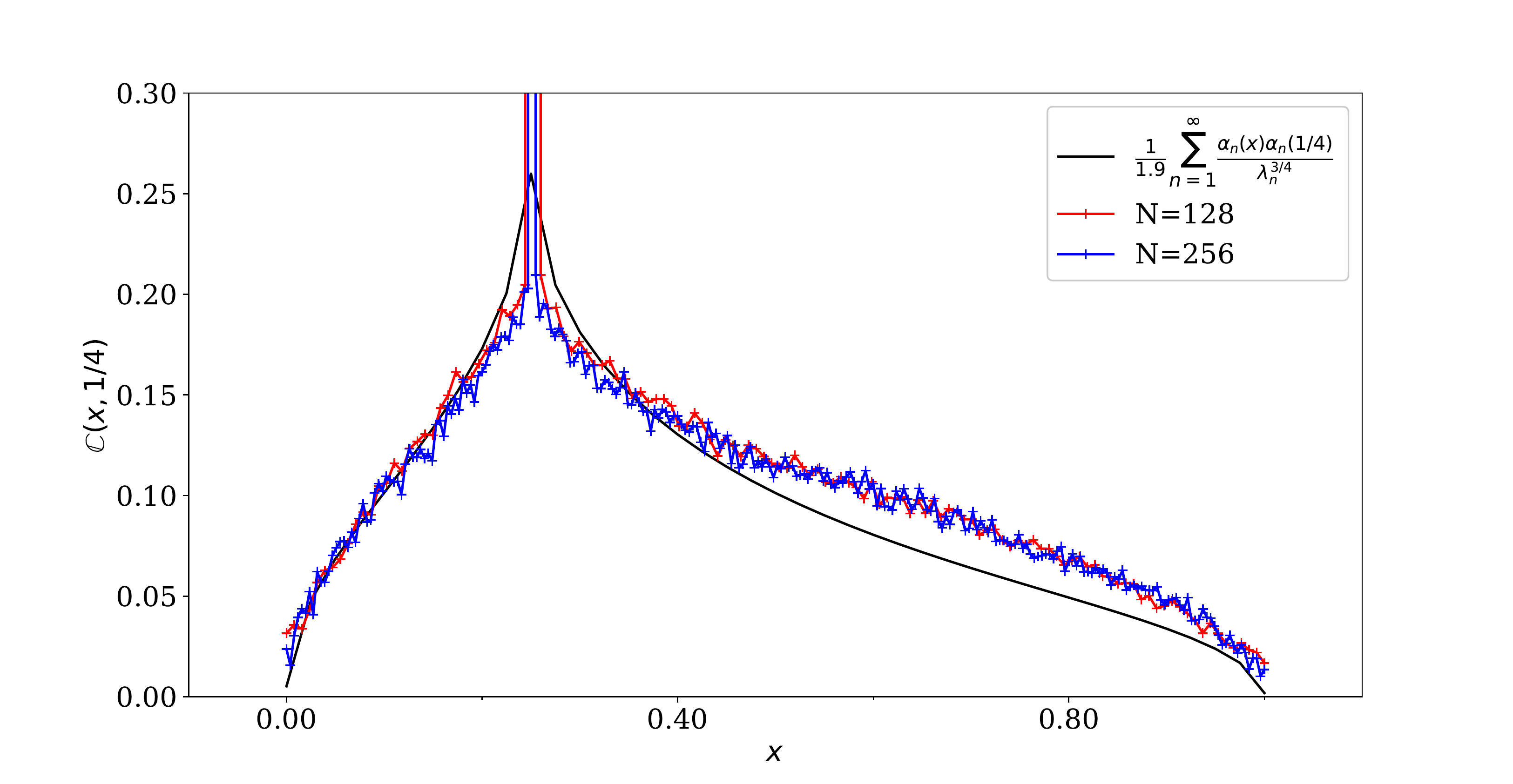}
			\includegraphics[width=0.65\linewidth]{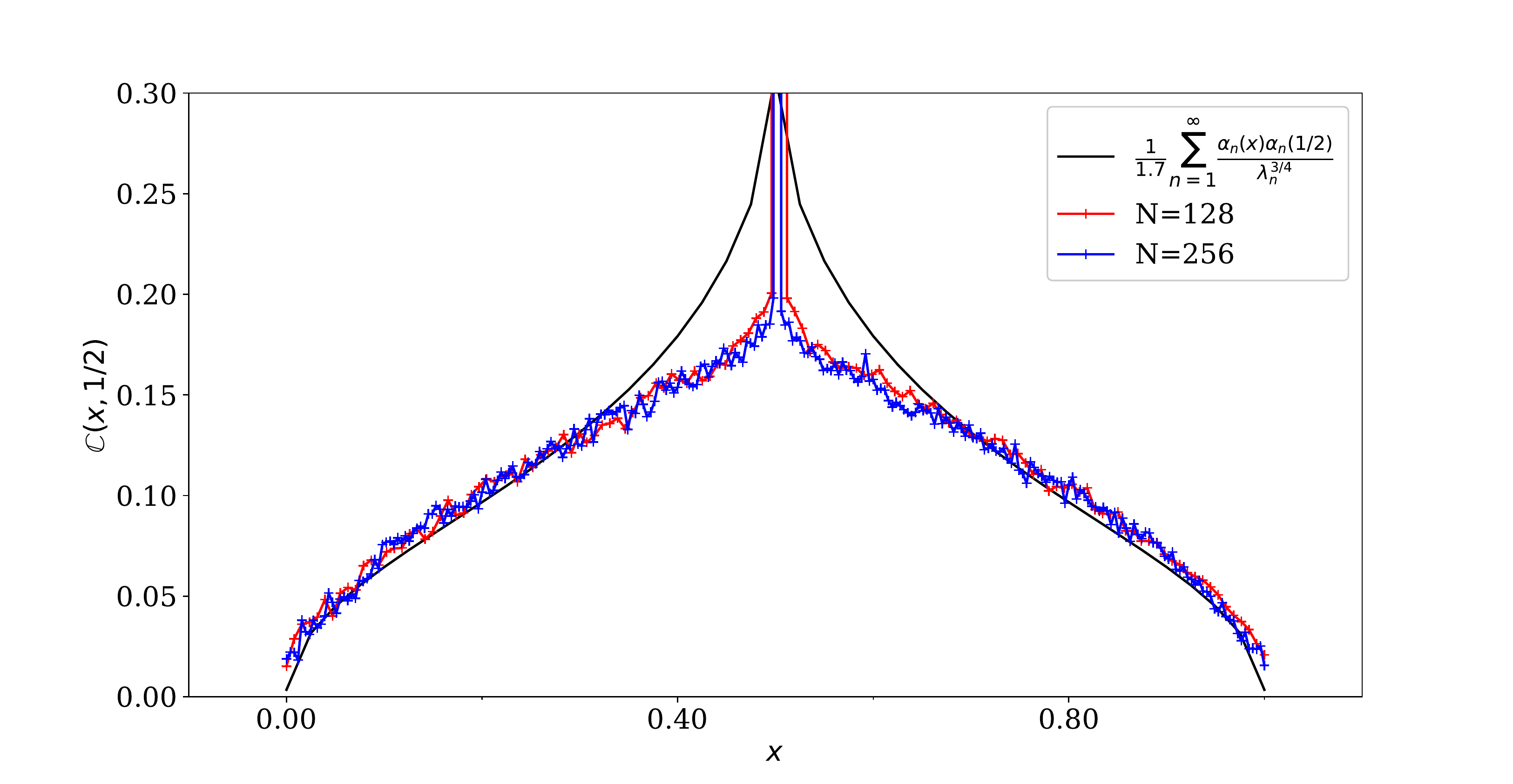}
			\caption{Here we show that instead of using $\psi_n$ and $\mu_n$ if we use $\sin$ and $\lambda_n^{3/4}$ for construction of inverse of the fractional operator, there are significant difference between the simulations with the formula $\mb{C}(x,y) = \sum_{n=1}^{\infty} \frac{\alpha_n(x)\alpha_n(y)}{\lambda_n^{3/4}}$}    
			\label{appfig:longrangesin}
		\end{figure}
	\end{document}